\documentclass[12pt]{article}
\usepackage[margin=2.5cm]{geometry}  
\usepackage{graphicx,psfrag,epsf}
\usepackage{enumerate}
\usepackage{url} 
\usepackage{siunitx}
\usepackage{float}
\usepackage{algorithm}
\usepackage{algorithmic}
\usepackage{amsmath}
\usepackage{amssymb}
\usepackage{graphicx}
\usepackage{subcaption}
\usepackage{natbib}
\usepackage{lineno}
\usepackage{dcolumn}
\usepackage{multirow}
\usepackage{comment}
\usepackage{booktabs}
\usepackage{colortbl}
\usepackage{rotating}
\usepackage{epstopdf}
\usepackage[table]{xcolor}  
\usepackage[utf8]{inputenc}
\newcolumntype{.}{D{.}{.}{-1}}
\newcolumntype{d}[1]{D{.}{.}{#1}}
\usepackage{hyperref}
\hypersetup{
	colorlinks=true,
	citecolor=blue,
}

\usepackage{rotating}
\usepackage{epsfig}%
\usepackage{epstopdf}

\def\bA{\mbox{\boldmath $A$}}

\def\bf{\mbox{\boldmath $f$}}

\def\1{\mbox{1}}

\def\bveps{\mbox{\boldmath $\varepsilon$}}

\def\bPhi{\mathbf{\Phi}}

\def\bveps{\mbox{\boldmath $\varepsilon$}}

\def\bPhi{\mathbf{\Phi}}

\def\bA{\mathbf{A}}

\def\bJ{\mathbf{J}}

\def\by{\mathbf{y}}

\def\0{\mbox{\bf{0}}}

\def\by{\mathbf{y}}

\def\0{\mbox{\bf{0}}}

\def\bJ{\mathbf{J}}

\def\by{\mathbf{y}}

\def\0{\mbox{\bf{0}}}

\newcommand{\nin}{\noindent}

\def\bkR{{\rm I\kern-.17em R}}

\def \1n{1\hskip -3pt \mbox{N}}

\newfont{\bbf}{cmbx12 scaled 1435}
\begin{document}
\setlength{\baselineskip}{.26in}
\thispagestyle{empty}
\renewcommand{\thefootnote}{\fnsymbol{footnote}}
\vspace*{0cm}
\begin{center}

\setlength{\baselineskip}{.32in}
{\bbf Regularized Generalized Covariance (RGCov) Estimator}\\

\vspace{0.5in}

\large{Francesco Giancaterini}\footnote{Universit`a di Roma "Tor Vergata", Italy, e-mail:{\it francesco.giancaterini@uniroma2.it}}, \large{Alain Hecq}\footnote{Maastricht University, The Netherlands, 
e-mail:{\it a.hecq@maastrichtuniversity.nl}}, \large{Joann Jasiak}\footnote{York University, Canada, 
e-mail:{\it jasiakj@yorku.ca}},\large{ Aryan Manafi Neyazi }\footnote{York University, Canada, e-mail: {\it aryanmn@yorku.ca } \\

The authors thank C. Gourieroux for helpful comments and acknowledge the financial support of the Natural Sciences and Engineering Council of Canada (NSERC) and Mathematics of Information Technology and Complex Systems (MITACS) of Canada.
}

\setlength{\baselineskip}{.26in}
\vspace{0.4in}

\today\\

\medskip

\vspace{0.3in}
\begin{minipage}[t]{12cm}
\small
\begin{center}
Abstract \\
\end{center}
We introduce a regularized Generalized Covariance (RGCov) estimator as an extension of the GCov estimator to high dimensional setting that results either from high-dimensional data or a large number of nonlinear transformations used in the objective function.
The approach relies on a ridge-type regularization for high-dimensional matrix inversion in the objective function of the GCov. 
The RGCov estimator is consistent and asymptotically normally distributed. We provide the conditions under which it can reach semiparametric efficiency and discuss the selection of the optimal regularization parameter. We also examine the diagonal GCov estimator, which simplifies the computation of the objective function. The GCov-based specification test, and the test for nonlinear serial dependence (NLSD) are extended to the regularized RGCov specification and RNLSD tests with asymptotic Chi-square distributions. Simulation studies show that the RGCov estimator and the regularized tests perform well in the high dimensional setting. We apply the RGCov to estimate the mixed causal and noncausal VAR model of stock prices of green energy companies. 

\bigskip

\end{minipage}

\end{center}
\renewcommand{\thefootnote}{\arabic{footnote}}

\newpage
\section{Introduction}

Mixed causal and noncausal processes have become increasingly popular for modeling nonlinear and dual patterns in time series, especially in financial contexts such as cryptocurrency rates and commodity prices. These financial variables often exhibit bubbles and other short-lasting nonlinear and explosive patterns [\cite{hencic2015noncausal}, \cite{lof2017noncausality},  \cite{gourieroux2017noncausal}, \cite{gourieroux2021forecast}]. In this context, it is crucial to recognize that conventional measures of past dependence, often referred to as "causal dependence," may not sufficiently capture the temporal dependence of these variables. This motivates interest in mixed causal and noncausal processes, as discussed by \cite{breid1991maximum}, \cite{lanne2011noncausal}, \cite{lanne2013noncausal}, \cite{gourieroux2016filtering}, \cite{hecq2016identification}, \cite{gourieroux2017local}  and \cite{cubadda2025sequential}.\\
\indent In the univariate framework, several studies investigated the nonlinear characteristics in financial data, especially bubbles, by using mixed causal and noncausal models [\cite{hencic2015noncausal}, \cite{fries2019mixed}, \cite{fries2019mixed}, \cite{gourieroux2021forecast}; \cite{gourieroux2021convolution}; \cite{hecq2021forecasting}]. Noncausal modeling techniques have also been applied to examine processes in a multivariate context (\cite{lanne2013noncausal},  \cite{gourieroux2017noncausal},\cite{cubadda2019detecting},  \cite{gourieroux2022nonlinear}, \cite{rygh2022causal}, \cite{cubadda2023detecting}, \cite{cubadda2024optimization} and \cite{cubadda2025sequential}). Most studies, if not all, however, only consider multivariate mixed causal-noncausal models for a small number of time series and for a relatively small number of nonlinear transformations in the objective function. This paper investigates both high dimensional issues and introduces the Regularized Generalized Covariance (RGCov) estimator, an extension of the Generalized Covariance (GCov) estimator introduced in \cite{gourieroux2017noncausal,gourieroux2023generalized} and further examined in \cite{jasiakneyazi} and \cite{cubadda2024optimization}. The GCov estimator is specifically designed for strictly stationary non-Gaussian processes, accommodating both causal and noncausal serial dependence. 

The Ridge type regularization is a natural approach for the GCov. Indeed, the GCov estimator is a Portmentau type test that requires the inversion of the variance-covariance matrix. This regularization makes the eigenvalues "bigger" and prevents the matrix becoming non-invertible. A Lasso regularization would put to 0 the small-valued eigenvalues on the covariance matrix and makes it non-invertible. Note that the causal-noncausal VAR parameters are not regularized in this paper and their estimations will be used to build portfolios that are for instance free from nonlinear bubble patterns. This is different to approaches in which VARs are sparsified in a high-dimensional setting (see e.g. Hecq et al. (2023) for a postdoble selection approach of causal VARs).



The GCov estimator is favored here to the maximum likelihood estimator to identify and estimate mixed models because it does not require any distributional assumption on the errors of the process, other than their non-Gaussianity.
It is a consistent, asymptotically normally distributed one-step estimator that is semiparametrically efficient. It can become parametrically efficient for well-selected nonlinear transformations.
The GCov estimator involves the autocovariances of nonlinear functions of the process that summarize nonlinear and noncausal serial dependence [\cite{chan2006note}]. It minimizes an objective function resembling a multivariate coefficient of determination, which includes the inverse of the variance matrix of a time series, i.e. its autocovariance at lag 0. When the dimension of the multivariate process is high and/or a high number of nonlinear functions are considered, this inversion may be difficult. This is the motivation for regularizing the variance in the formula of the GCov estimator. 

The rest of the paper is as follows. In Section 2 we recap the main results about the GCov estimator. Section 3 proposes our new RGCov estimator whose behavior is investigated in Section 4 with a Monte Carlo study. Section 5 estimates a mixed causal-noncausal VAR for stocks that belong to the Rennix green index. Section 6 concludes. 

\setcounter{equation}{0}\def\theequation{2.\arabic{equation}}

\section{The VAR Model and GCov Estimator}

This section describes the causal-noncausal Vector Autoregressive (VAR) model and the semi-parametric GCov estimator along with the associated specification test.

\subsection{The Model}

Let us consider a strictly stationary process $\{Y_t\}$ of dimension $n$ satisfying a semi-parametric model:
\begin{equation}
\label{3.1}
g(\tilde{Y}_t; \theta) = u_t,
\end{equation}

\nin where $g$ is a known function,
$\tilde{Y}_t =(Y_t, Y_{t-1},...,Y_{T-p})$, $p$ is an integer,
$(u_t)$ is an i.i.d. sequence, and $\theta$ is an unknown parameter vector. We assume that the model is well specified and the true value of the parameter $\theta$ is $\theta_0$. 

\medskip
An example of process (2.1) is the causal-noncausal VAR($p$) model. The multivariate causal-noncausal VAR($p$) process is:

$$Y_t = \Phi_1 Y_{t-1}+ \cdots + \Phi_p Y_{t-p} + u_t,$$

\nin where $\theta = [vec \Phi_1', ..., vec \Phi_p']'$ \footnote{For any $m \times n$ matrix $A$ whose $j$th column is $a_j, \; j= 1, . . ., n$, $vec(A)$ denotes the column vector of dimension $mn$ defined as $vec(A) = (a_1'...,a_j',...,a_n')',$
where the prime denotes transposition.} and error $u_t$ is a multivariate non-Gaussian i.i.d. process with finite fourth-order moments. 
The roots of the characteristic equation $det(Id - \Phi_1 z - \cdots  \Phi_p z^p) = 0$
are of a modulus either strictly greater than or smaller than one. Then, there exists a unique (strictly) stationary solution $(Y_t)$ with a two-sided representation $MA(\infty)$, which satisfies model (\ref{3.1}) with:

\centerline{$g(\tilde{Y}_t, \theta)  =  Y_t - \Phi_1 Y_{t-1} - \cdots  - \Phi_p Y_{t-p} = u_t(\theta).$}

\nin The causal-noncausal VAR($p$) model has been studied in \cite{gourieroux2016filtering,gourieroux2017noncausal}, \cite{davis2020noncausal}, \cite{rygh2022causal}, \cite{cubadda2024optimization}, and \cite{hall2024modelling}. The error $u_t=u_t(\theta_0)$ cannot be interpreted as an innovation, because it is correlated with the past $y's$. Moreover, even though the function $g$ is linear in the current and lagged values of $Y_t$, the presence of noncausal components in $Y_t$ implies nonlinear dynamics of $Y_t$ from the calendar time perspective, with $E(Y_t|\underline{Y_{t-1}})$ nonlinear in $\underline{Y_{t-1}} = (Y_t, Y_{t-1},...)$ and  conditional heteroscedasticity $V(Y_t| \underline{Y_{t-1}})$.

The presence of noncausal serial dependence can be detected through the analysis of nonlinear autocovariances, i.e. autocovariances of nonlinear functions of the observed process. This approach follows from \cite{chan2006note}, who show that the presence of nonlinear dependence in strictly stationary non-Gaussian time series is revealed by the autocovariances of nonlinear transforms of that time series.

Let us consider nonlinear functions transforming a multivariate process $g(y_t;\theta)$ of dimension $n$ into a multivariate process $v_t(\theta)$ of dimension $K=Jn$ with the components $a_j[g_i(y_t;\theta)], j=1,...,J, i=1,...,n$. The transformed process:

\begin{equation}
\label{3.2}
v_t(\theta) = \left( \begin{array}{c}
     a_1[g_1(\tilde{y}_{t}; \theta)]  \\
      \vdots \\
    a_J[g_n(\tilde{y}_{t};\theta)] \\
    \vdots \\
    \vdots \\
     a_1[g_1(\tilde{y}_{t};\theta)]  \\
     \vdots \\
     a_J[g_n(\tilde{y}_{t}; \theta)] 
\end{array}  \right),
\end{equation}

\nin is also serially i.i.d. when $\theta=\theta_0$.  The transformed process has a dimension higher than $n$ because it is augmented by nonlinear differentiable functions of $g(Y_t;\theta)$, such as squares or logarithms, for example. Specifically, if $g (Y_t;\theta)$ has no finite fourth-order moment, then it can be replaced by a transformed multivariate process $v_t$ with a finite fourth-order moment to ensure the validity of the estimation procedure.

\subsection{The GCov estimator}

The advantage of the semi-parametric Generalized Covariance (GCov) estimator introduced in \cite{gourieroux2017noncausal,gourieroux2023generalized}  is that it does not require any distributional assumptions on the true errors $u_t=u_t(\theta_0)$, other than being i.i.d. and non-Gaussian. In addition, it is easy to compute and leads to a specification test with a known limiting distribution.

\subsubsection{The Semi-parametrically efficient GCov}
The GCov estimator of $\theta$ in model (\ref{3.1}) is defined as:
\begin{equation}
\hat{\theta}_T = Arg min_{\theta} \sum_{h=1}^H Tr \hat{R}_T^2(h; \theta),
\label{eq:GCov}
\end{equation}

\nin where
\begin{equation}
\hat{R}_T^2(h ; \theta)=\hat{\Gamma}_T(h; \theta) \hat{\Gamma}_T(0;\theta)^{-1}\hat{\Gamma}_T(h;\theta)' \hat{\Gamma}_T(0;\theta)^{-1}.
\end{equation}

\nin and $\hat{\Gamma}_T(h;\theta)$ is the sample autocovariance between $v_t(\theta)$ and $v_{t-h}(\theta)$, with $h$ denoting the lag. For a sample of $T$ observations $Y_1,...,Y_T$, the sample autocovariances are obtained from the process $\{ v_t(\theta) \}, t=1,...,T$ of dimension $K=Jn$:

\begin{equation}
\label{Gamma}
   \hat{\Gamma}_T(h; \theta) = \frac{1}{T} \sum_{t=h}^T  v_t(\theta) v_{t-h}(\theta)^{'} - \frac{1}{T} \sum_{t=h}^{T-1} v_t(\theta)^{'} \frac{1}{T} \sum_{t=h+1}^T  v_{t-h}(\theta).
\end{equation}

\nin for $h=0,...,H$.

The GCov estimator minimizes an objective function, which is equivalent to the sum of sample multivariate coefficients of determination computed at different lags $h$ from nonlinear transforms $v_t(\theta)$. Under the regularity conditions given in Appendix A, the GCov estimator is consistent, asymptotically normally distributed, and semi-parametrically efficient. It can achieve parametric efficiency for well-selected transformations, given in \cite{gourieroux2023generalized}.

The GCov objective function can be used to test the goodness of fit of the model by testing the absence of nonlinear and linear serial correlation in the residuals: $H_0:$ $\Gamma(h) = 0$ for $h=1,2....$. The test statistic $\hat{\xi}_T^a(H) =  T \sum_{h=1}^H Tr \hat{R}_T^2(h; \hat{\theta}_T)$ follows asymptotically a $\chi^2(K^2 H - (dim \theta))$ distribution where $K=Jn$, when the process $\{Y_t\}$ satisfies $g(\tilde{Y}_t; \theta_0) =u_t$ and $u_t$ is serially i.i.d.. The goodness of fit test at level $\alpha$ is conducted as follows: the null hypothesis $H_0$ is rejected when $\hat{\xi}_T (H) > \chi^2 _{1-\alpha}(K^2H - dim(\theta))$ and $H_0$ is not rejected otherwise.

\medskip
 For preliminary data analysis, the NLSD test of the absence of nonlinear and linear serial dependence in time series introduced in [\cite{jasiakneyazi}] can be used.
 It is inspired by the specification test described above.
 The NLSD test needs to be applied prior to the estimation to determine whether there is evidence of noncausal serial dependence in the data. Thus, it corresponds to the special case of a strong white noise $Y_t=u_t$, where there is no parameter $\theta$ to be estimated.
 The test statistic is computed directly from the data transforms $v_t = [a_1(Y_{1,t}),...,a_1(Y_{n,t})....,a_1(Y_{1,t}),...,a_J(Y_{n,t})]'$ to test the null hypothesis of the absence of nonlinear and linear dependence based on the transformed time series of dimension $Jn=K$. It is based on the statistic:
$\hat{\xi}_T(H) =  T \sum_{h=1}^H Tr \hat{R}_T^2(h)$, where:

$$
\hat{R}_T^2(h)=\hat{\Gamma}_T(h) \hat{\Gamma}_T(0)^{-1} \hat{\Gamma}_T(h)' \hat{\Gamma}_T(0)^{-1}.
$$

\nin where  

\begin{equation}
\hat{\Gamma}_T(h) = \frac{1}{T} \sum_{t=h}^T  v_t v_{t-h}^{'} - \frac{1}{T} \sum_{t=h}^{T-1} v_t \frac{1}{T} \sum_{t=h+1}^T  v'_{t-h}.
\end{equation}

\nin for $h=0,...,H$. Under the serial independence of $\{Y_t\}$, this statistic follows asymptotically a $\chi^2(K^2 H)$ distribution.

\subsubsection{Diagonal GCov}

The diagonal GCov estimator $\hat{\theta}_T^d$ is obtained by minimizing the objective function with the variance matrix $\hat{\Gamma}_T^d(0; \theta)$ containing only diagonal elements of $\hat{\Gamma}_T(0; \theta)$. This estimator is described by \cite{gourieroux2017noncausal} [see also \cite{cubadda2011testing}] and is given by:

\begin{equation}
\hat{\theta}_T^d = Arg min_{\theta} \sum_{h=1}^H Tr \hat{R}^2_{d,T} (h; \theta),
\label{eq:GCov17}
\end{equation}

\nin where
\begin{equation}
\hat{R}^2_{d,T} (h ; \theta)=\hat{\Gamma}_T(h; \theta) \hat{\Gamma}_T^d(0;\theta)^{-1}\hat{\Gamma}_T(h;\theta)' \hat{\Gamma}_T^d(0;\theta)^{-1},
\end{equation}

\nin with the matrix $\hat{\Gamma}_T^d(0;\theta)$ being a diagonal matrix containing only the variances of $v_{i,t}(\theta), i=1,...,K$.

This estimator is not semiparametrically efficient, as it is not optimally weighted. Its asymptotic properties are described in \cite{gourieroux2017noncausal}.

\subsection{Nonlinear Transformations}
One may choose a high number of nonlinear transformations to improve the efficiency of the GCov estimator and asymptotic performance of the GCov specification test. 
The additional nonlinear transformation increase the dimension of matrix $\hat{\Gamma}_T(0; \theta)$, and inverting the variance matrix of a large dimension can become numerically challenging. \cite{jasiakneyazi} discuss how  to select the basis of transformations that can provide more information on the parameters, in addition to the linear and quadratic functions of errors $u_t$.
In the causal-noncausal models with extreme risks and local explosive patterns, including the bubbles, the error does not necessarily has power moments. Moreover, some of the parameters driving those extreme risks, maybe non-identifiable
from the transformations $u_t, u_t^2$ only. The following system of generators can be considered:

$$\mathcal{A} = \{ a_{t,p}(u) = |u|^p \exp(-t |u|), \; p \in \mathbb{N}, t \in [0,1] \},$$

\nin with a countable dense subsystem given by:

$$\mathcal{A}_n = \{ a_{t_{j,n},p}(u) = |u|^p \exp(-t_{j,n}|u|), \; p \in \mathbb{N}, t_{j,n} \in [0,1], \; j=1,...,n \},$$

\nin where $(t_{1,n},...,t_{n,n})$ is dense in $[0,1]$ when $n$ tends to infinity [see, \cite{bierens1990consistent}, page 1448, for a similar approach].
The decreasing exponential functions provide square integrability of the power transforms ensuring adequate weighting. 

\setcounter{equation}{0}\def\theequation{3.\arabic{equation}}
\section{The RGCov Estimator}

When $n$ is high but finite, the $ K \times K$ matrix $\hat{\Gamma}_T(0; \theta)$
is of high dimension, and its inverse may be difficult to compute numerically.
This is, for example, the case of a causal-noncausal VAR process with a large number $n$ of components. In general, the dimension of the parameter $\theta=[vec\Phi_1',...,vec\Phi'_p]'$ of the VAR($p$) process is $p \times n^2$. Therefore, it is necessary to use non-linear transformations $J \geq pn$ to ensure the identifiability of the model. In the particular case of a VAR process, a large lag $p$ increases the dimension too, requiring in turn a large number of nonlinear transformations $J$.
Therefore, the three parameters $n,J$ and $p$ play an important role in determining the dimension of the variance matrix. There are two issues with inverting the matrix $\hat{\Gamma}_T(0,\theta)$.

\nin (i) The spectral decomposition of this matrix leads to close to zero smallest eigenvalues. Then, the inversion problem is in the numerical inversion given these close-to-zero eigenvalues. We refer to it as an issue of weak invertibility. 

\nin (ii) Moreover, even when the eigenvalues are sufficiently different from zero and positive, the available inversion algorithms may not be numerically efficient.

The first issue can be solved by introducing a regularized objective function, and the second one by applying an algorithm that updates the inverse of the regularized variance-covariance matrix with each consecutive observation.

\subsection{Definition}
\label{subsec:def}
To solve the problem of weak invertibility of matrix $\hat{\Gamma}_T(0; \theta)$, we introduce the Regularized (RGCov) estimator.

\medskip
\nin \textbf{Definition 1:}
\nin i) A Regularized (RGCov) estimator of $\theta_0$ in model (\ref{3.1}) is defined as:
\begin{equation}
\hat{\theta}_{T}(\delta) = Arg min_{\theta} L_T(\theta, \delta),
\label{eq:RGCov}
\end{equation}

\nin where the objective function:

$$
L_T(\theta, \delta_T) = \sum_{h=1}^H Tr \hat{R}_T^2(h; \theta, \delta),
$$

\nin with

\begin{equation}
\hat{R}_T^2(h ; \theta, \delta)=\hat{\Gamma}_T(h; \theta) \hat{\Gamma}_T(0, \theta, \delta)^{-1}   \hat{\Gamma}_T(h;\theta)' \hat{\Gamma}_T(0, \theta, \delta)^{-1},
\end{equation}

\nin depends on the regularized variance matrix:

\begin{equation}
\hat{\Gamma}_T(0, \theta, \delta) =  \delta I+ \hat{\Gamma}_T(0;\theta),
\end{equation}

\nin where $I$ is the identity matrix of dimension $K$ and $\delta>0$ is the shrinkage coefficient. 

\medskip
\nin ii) More generally, a RGCov can be defined as $\hat{\theta}_T =\hat{\theta}_T(\delta_T)$, where the shrinkage parameter $\delta_T, \delta_T>0$, tends to zero, or to $\delta>0$, when $T$ tends to infinity.
This includes as a special case $\delta_T=\delta>0, \forall T$, considered in i).

\medskip

The regularization with fixed $\delta>0$ is used as in a ridge regression to solve the weak invertibility issue, but it leads to an estimator that is not asymptotically efficient. The asymptotic efficiency is recovered by adjusting appropriately the shrinkage parameter when the number of observations $T$ increases. 

\subsection{The Asymptotic Properties}

\subsubsection{Asymptotic Properties of RGCov Estimator}

\nin Let us consider first the general case when the sequence of positive shrinkage parameters $\delta_T \rightarrow \delta \geq 0$, and then the case when  $\delta_T \rightarrow 0$.

\nin a) In the general case, the objective function is

$$L_T(\theta, \delta_T) = \sum_{h=1}^H Tr[\hat{\Gamma}_T(h, \theta) \hat{\Gamma}_T(0, \theta, \delta_T)^{-1} \hat{\Gamma}_T(h, \theta) \hat{\Gamma}_T(0, \theta, \delta_T)^{-1}].$$

\textbf{Proposition 1}:

\nin Under the regularity conditions given in Appendix A:

i) $\hat{\theta}_T=\hat{\theta}_T(\delta_T)$ is consistent of $\theta_0$

ii) 
$\sqrt{T} (\hat{\theta}_T  - \theta_0)$ 
$ \stackrel{d}{\rightarrow} N(0, J(\theta_0, \delta)^{-1} I(\theta_0, \delta) J(\theta_0, \delta)^{-1})$, where matrices $J(\theta_0, \delta), I(\theta_0, \delta)$ are:

$$
J (\theta_0, \delta)  =  2 \sum_{h=1}^H \left\{ \frac{\partial vec \Gamma(h; \theta_0)'}{\partial \theta} [\Gamma (0; \theta_0, \delta)^{-1}  \otimes   
\Gamma (0; \theta_0, \delta)^{-1} ] \frac{\partial vec \Gamma(h; \theta_0)}{\partial \theta'} \right\},
$$

\nin and

\begin{eqnarray*}
I(\theta_0, \delta) & = & 4\sum_{h=1}^H \left\{ \frac{\partial vec \Gamma(h; \theta)'}{\partial \theta} \left[\Gamma(0; \theta_0, \delta)^{-1} \otimes \Gamma(0; \theta_0, \delta)^{-1}\right] [\Gamma(0;\theta_0) \otimes \Gamma(0;\theta_0)] \nonumber \right. \\
& & \left. \left[\Gamma(0; \theta_0, \delta)^{-1} \otimes \Gamma(0; \theta_0, \delta)^{-1}\right]  \frac{\partial vec \Gamma(h; \theta)}{\partial \theta'} \right\}.
\end{eqnarray*}

\nin Thus the speed of convergence of the estimator and its asymptotic distribution do not depend on the way $\delta_T$ tends to $\delta$. However, its asymptotic variance-covariance matrix depends on $\delta$.

Because this estimator is not optimally weighted, it is not semi-parametrically efficient and belongs in the class of covariance estimators considered in \cite{gourieroux2017noncausal} [see also \cite{cubadda2011testing}]. 

\medskip

\nin b)  In the special case $\delta=0$, the expression of the asymptotic distribution is simplified.

\medskip
\textbf{Proposition 2}:

\nin Under the regularity conditions given in Appendix A, if $\delta_T \rightarrow 0$ when $T \rightarrow \infty$:

i) $\hat{\theta}_T = \hat{\theta}_T(\delta_T)$ is consistent of $\theta_0$.

ii) $\sqrt{T} (\hat{\theta}_T  - \theta_0) \stackrel{d}{\rightarrow} N(0, J(\theta_0)^{-1})$, where $J(\theta_0)= J(\theta_0,0)/2$ and

$$
J (\theta_0, \delta)/2 = I(\theta_0, 0)/4 =  \sum_{h=1}^H \left\{ \frac{\partial vec \Gamma(h; \theta_0)'}{\partial \theta} [\Gamma (0; \theta_0)^{-1}  \otimes   
\Gamma (0; \theta_0)^{-1} ] \frac{\partial vec \Gamma(h; \theta_0)}{\partial \theta'} \right\}.
$$

\nin where $\Gamma (0; \theta_0) = \Gamma (0; \theta_0, 0)$.

\medskip

\nin Since the asymptotic properties do not depend on the way $\delta_T$ tends to 0, these properties are the same as the properties of the GCov estimator. In particular, it is asymptotically semi-parametrically efficient [\cite{gourieroux2023generalized}].

\medskip
The different expressions of $J (\theta_0, \delta)$ and $I (\theta_0, \delta)$ (up to factor 4) are similar to the expressions appearing in the variance of a Generalized Least Squares (GLS) estimator in a regression model with a weighting matrix $\Omega^{-1}$ when the true weights are $\Omega_0$, say, that is:

$$(X'\Omega^{-1}X)^{-1} (X' \Omega^{-1} \Omega_0 \Omega^{-1} X) (X' \Omega^{-1} X)^{-1},$$

\nin where $X = \frac{\partial vec \Gamma'(h, \theta_0)}{\partial \theta}$, $\Omega = \Gamma(0, \theta_0, \delta) \otimes \Gamma(0, \theta_0, \delta) $, and
$\Omega_0 = \Gamma(0, \theta_0) \otimes \Gamma(0, \theta_0) $.

\subsubsection{Asymptotic Properties of the RGCov and RNLSD Test Statistics }
The regularization with $\delta_T$ can be used to build test statistics for testing the model specification, or the data for the absence of linear and nonlinear dependence.

 We can use the regularized estimator $\hat{\theta}_T=\hat{\theta}_T(\delta_T)$ in the formulas of test statistics for testing the fit of the model to extend the GCov test introduced in [Gourieroux, Jasiak (2023)]). Let us define the residual-based Regularized GCov (RGCov) test statistic for testing the model specification: 

\begin{equation}
\hat{\xi}_T(H, \delta_T) =  T \sum_{h=1}^H Tr \hat{R}_T^2(h, \hat{\theta}_T(\delta_T), \delta_T).
\end{equation}

\nin Let us again first  consider the general case.

\medskip
\textbf{Proposition 3:}
\nin i) If $\delta_T \rightarrow \delta \geq 0$, when $T \rightarrow \infty$, then under the null hypothesis of independence of $u_t's$, the RGCov test statistic is asymptotically distributed as a positive combination of independent chi-square variables:

$$ \hat{\xi}_T(H, \delta_T)  \stackrel{d}{\rightarrow} \sum_{l=1}^L   \lambda_{l} Z_{l},$$

\nin where $L=K^2H - dim \theta$, $Z_l, \, l=1,...,L$ are independent $Z_l \sim \chi^2(1)$ and $\lambda_{l}>0, \, \forall l$.

ii) In particular, when $dim \theta = 0$, i.e. for the Regularized NLSD (RNLSD) test statistic, we have:

$$ \hat{\xi}_T(H, \delta_T)  \stackrel{d}{\rightarrow} \sum_{l=1}^{K^2}   \lambda_{l}^* Z_{l}^*,$$

\nin where $Z_l^*, \, l=1,...,K^2$ are independent $\chi^2(H)$ variables and $\lambda_{l}^*, l=1,...,K^2$ are the eigenvalues of the symmetric positive definite matrix:

$$[\Gamma(0)^{-1/2} \Gamma(0, \delta) \Gamma(0)^{-1/2}] \otimes [\Gamma(0)^{-1/2} \Gamma(0, \delta) \Gamma(0)^{-1/2}].$$

\nin The properties of the tensor (Kronecker) product imply that  $\lambda_{l}^*$ are obtained by considering all the products $\mu_j \mu_k$, where $\mu_k, \, k=1,...,K$ are the eigenvalues of $\Gamma(0)^{-1/2} \Gamma(0, \delta) \Gamma(0)^{-1/2}$.

Proof: See Appendix A.

\medskip
When $dim(\theta)>0$ the test statistic (3.4) is testing the specification of the semi-parametric model, and it is applied to the residuals of the model estimated by the RGCov estimator. The case $dim(\theta)=0$ corresponds to the regularized extension of the NLSD test [\cite{jasiakneyazi}].
The Regularized NLSD (RNLSD) test statistic tests the null hypothesis of the absence of linear and nonlinear dependence in the data themselves:

$$H_{0} = (\Gamma (h) = 0, \; h=1,...,H),$$
\medskip
\nin In the special case $\delta=0$, the test statistic (3.4) has a chi-square asymptotic distribution.

\medskip
\textbf{Proposition 4:}
\nin If $\delta_T \rightarrow 0$ when $T \rightarrow \infty$, then it follows from Proposition 3 and the asymptotic equivalence of the GCov and RGCov estimators that, under the null hypothesis of independence of $u_t's$,  the RGCov test statistic (3.4) has asymptotically a chi-square distribution with the degree of freedom equal to $K^2H-dim(\theta)$, for $dim(\theta) \geq 0$.

\medskip
When the degree of freedom is greater than 30, than under the null hypothesis of independence of $u_t's$, the following function of the test statistic (3.4) evaluated for a given $H$ and $\delta$, and of the degree of freedom denoted by $\nu = K^2H-dim(\theta)$ with $dim(\theta) \geq 0$:

$$ \zeta(\hat{\xi}_T, \nu) = \sqrt{2 \, \hat{\xi}_T} - \sqrt{2 \nu -1 },$$

\nin is asymptotically normally distributed:

$$ \zeta(\hat{\xi}_T, \nu) \sim N(0,1).$$

\nin Hence, the null hypothesis can be alternatively tested using the function $\zeta$ and the asymptotically valid critical values of  standard normal. This approach is particularly useful for models of large dimension $n$, or when a high number of nonlinear transformations $J$ is considered. 

\subsection{Efficient Inverse Updating}

The implementation of the RGCov approach requires the inversion of matrices of high dimension, equal either to the total number $K$ of moments when computing $\hat{\Gamma}_T (0, \theta, \delta)$, or the number of parameters, when estimating the asymptotic variance-covariance matrix of the estimators. For numerical efficiency, the algorithms based on the Sherman-Morrison formula [\cite{sherman1949adjustment,sherman1950adjustment}] can be used\footnote{The approach based on the Sherman-Morrison formula is cheaper than the computation of the inverse from either a spectral decomposition, or a Cholesky decomposition. Note also that these decompositions are not unique.}.

We first recall this formula and next explain its implementation.

\medskip

\nin \textbf{Lemma 1} [\cite{sherman1949adjustment}, \cite{sherman1950adjustment}]: If $A$ is a symmetric positive definite matrix, then

$$(A+ xx')^{-1} = A^{-1} - \frac{A^{-1} x x' A^{-1}}{1+ x'A^{-1}x} $$

\nin Proof: We have 

$(A+xx')( A^{-1} - \frac{A^{-1} x x' A^{-1}}{1+ x'A^{-1}x})  = AA^{-1} + xx' A^{-1}(1 -  \frac{1}{1+x'A^{-1}x} - \frac{x' A^{-1}x}{1+x'A^{-1}x}) = I.$

\nin The result follows. \hfill QED

\medskip

The above Lemma can be used to compute recursively the 
matrix $C_T=[\rho_1 I + \rho_2 \sum_{t=1}^T x_t x_t']^{-1}$ from the sample of $t=1,...,T$ observations.

\medskip
\textbf{Corollary 1}: 
$$C_T = C_{T-1} - \frac{\rho_2 C_{T-1} x_Tx_T' C_{T-1}}{1+ \rho_2 x_T' C_{T-1} x_T}, \; T \geq 2.$$

\nin Proof: We can apply Lemma 1 with $A=\rho_1 I + \rho_2 \sum_{t=1}^{T-1} x_t x_t'$ and $\sqrt{\rho_2} x_t$ for $x$. The result follows by observing that $A^{-1} = C_{T-1}$. \hfill QED

\medskip

\nin The starting value $C_1$ in this recursion is given in the next corollary:

\medskip

\textbf{Corollary 2}:

$$C_1 = (\rho_1 I + \rho_2 x_1 x_1')^{-1} = \frac{1}{\rho_1} I - \frac{\rho_2}{\rho_1^2}
\frac{x_1x_1'}{ 1+ \frac{\rho_2}{\rho_1} x_1'x_1}.$$

Proof: The result follows by Lemma 1 applied  with $A=\rho_1I$ and $x= \sqrt{\rho_2} x_1$. \hfill QED

\medskip

\nin The recursive approach given in the above corollaries can be used to invert the matrix:

$$\hat{\Gamma}_T(0, \theta, \delta_T) = \delta_T I + \frac{1}{T} \sum_{t=1}^T [v_t(\theta) - \frac{1}{T}\sum_{t=1}^T v_t(\theta)][v_t(\theta) - \frac{1}{T}\sum_{t=1}^T v_t(\theta)]',
$$

\nin for given values of $T, \delta_T$ and $\theta$. The recursive formulas are then applied with $\rho_1=\delta_T$, $\rho_2 = \frac{1}{T}$, and $x_t = v_t(\theta)- \frac{1}{T}\sum_{t=1}^T v_t(\theta)$.

\medskip
This inversion approach can be used when the objective function is maximized by an algorithm that requires evaluating the value of the objective function at $\hat{\theta}_T^{(j)}$, where $\hat{\theta}_T^{(j)}$
is the approximation of $\theta$ at iteration $j$ of the algorithm.

\medskip

\nin  The above corollaries can also be used to compute numerically the asymptotically efficient variance-covariance matrix of the RGCov estimator when the inverse of the Hessian matrix is approximated by the inverse of the outer product of scores.  

\setcounter{equation}{0}\def\theequation{4.\arabic{equation}}
\section{Monte Carlo Studies}
This section evaluates the performance of the RGCov, GCov, and diagonal GCov estimators through three simulation studies, focusing on scenarios where the $K\times K$ matrix $\Gamma(0)$ is of high dimension. Specifically, in the first simulation study (Section~\ref{subsec:large_n}), the high dimensionality arises from the number of variables: we set $K=30$, generated with $n=15$ and $J=2$. In the second simulation study (Section~\ref{subsec:large_J}), the source of high dimensionality shifts from the number of variables to the number of linear and nonlinear transformations. In this regard, we again set $K=30$, but generated here with $n=3$ and $J=10$. Finally, in the third simulation study \ldots [continue with description].









\subsection{High Dimensionality Driven by the Number of Variables}
\label{subsec:large_n}
We evaluate here the performance of the three estimators mentioned above in scenarios where the high dimensionality of $\Gamma(0,\theta)$ is driven by $n$, the number of variables. It should be noted that the performance of the RGCov estimator is here examined under the two shrinkage settings: $\delta_T = \delta$ for the case $\delta_T \to \delta$ as $T$ increases, and $\delta_T = \eta/T$ when considering the special case $\delta_T \to 0$ (see Section \ref{subsec:def}).

The data-generating process (DGP) is a 15-dimensional mixed VAR(1) model consisting of 14 real eigenvalues inside the unit circle, $\lambda_{1}$=$\{$-0.437, -0.374, -0.360, -0.263, -0.248, -0.201, -0.162, -0.105, -0.004, 0.320, 0.277, 0.200, 0.162, 0.164$\}$, and one real eigenvalue outside the unit circle, $\lambda_{2} = 1.5$. The error term is assumed to follow a multivariate Student-$t$ distribution with degrees of freedom $\nu=4$ and a scale matrix equal to the identity matrix.

For all three estimators, we set $H=2$, $J=2$ and define the two transformations as follows:
\begin{equation}
    \forall\, i \in \{1, \ldots, n\}, \quad \forall\, t \in \{1, \ldots, T\}, \quad
\begin{cases}
a_1\bigl(g_i(y_t; \theta)\bigr) = u_{i,t},\\
a_2\bigl(g_i(y_t; \theta)\bigr) = u_{i,t}^2
\end{cases}.
\label{eq:transf}
\end{equation}
By combining the linear function $a_1$ with the quadratic function $a_2$, the estimator is well-equipped to capture both direct dynamics (first moment) and volatility dynamics (second moment) in the underlying data. Hence, $K = nJ = 30$ implies that matrices $\Gamma(h, \theta)$, for each $h \in \{0,1,2\}$, are of size $30 \times 30$. Therefore, inverting $\Gamma(0,\theta)$ for the GCov estimator is computationally challenging due to the curse of dimensionality. To address this issue, we investigate whether regularizing this matrix (RGCov) or considering only its diagonal elements (diagonal GCov) can lead to meaningful improvements.

\begin{figure}[H]
    \centering
    \caption{Bias, variance, and MSE of the (R)GCov estimator (first row) and the diagonal GCov estimator (second row) under the setting $\delta_T = \delta$, for varying values of $\delta$ and sample sizes $T$. Increasing values of $\delta$ are considered to facilitate cross-validation and identify the optimal shrinkage level that minimizes the respective performance metric. Each metric is computed element-wise and then averaged over all matrix entries. The high dimensionality in this setting arises from the number of variables $n$.}
    \includegraphics[width=18.5cm, height=6.7cm]{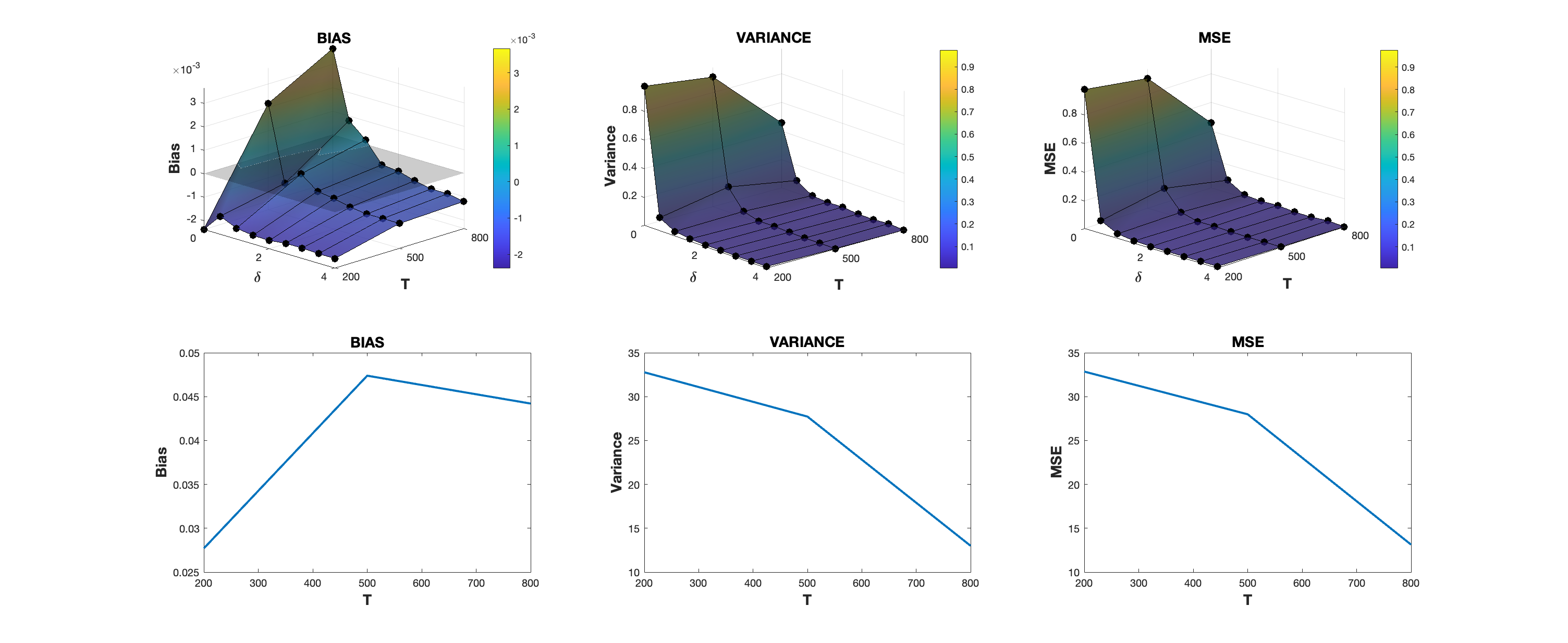}
    \label{fig:bias_variance_mseA}
\end{figure}

\begin{figure}[H]
    \centering
    \caption{The graph shows the bias, variance, and MSE of the RGCov estimator under the setting $\delta_T = \eta/T$, for increasing values of $\eta$ and sample sizes $T$. Each metric is computed element-wise and then averaged over all matrix entries. The high dimensionality in this setting arises from the number of variables $n$.}
    \includegraphics[width=18.5cm, height=3.3cm]{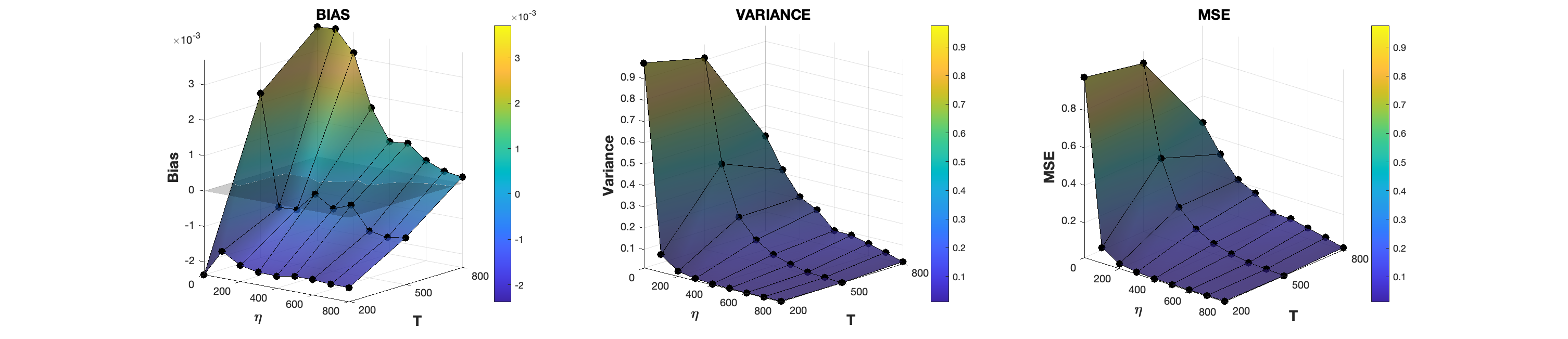}
    \label{fig:bias_variance_mseB_ss1}
\end{figure}

We assess the performance of the three estimators by analyzing: bias, estimated variance and mean squared error (MSE). In particular, to measure the bias of the coefficient in the $i$-th row and $j$-th column of the autoregressive matrix, we use the following expression:
\begin{equation} 
\label{eq:Bias}
    \widehat{\text{Bias}}(\hat{\Phi}_{ij}) = \frac{1}{M} \sum_{m=1}^{M} 
    \left( \hat{\Phi}^{(m)}_{ij} - \Phi_{ij,0} \right) = \frac{1}{M} 
    \sum_{m=1}^{M} \hat{\Phi}^{(m)}_{ij} - \Phi_{ij,0} = \overline{\Phi}_{ij}- \Phi_{ij,0}.
\end{equation} 
where $i, j = 1, \dots, n$; $\hat{\Phi}^{(m)}_{ij}$ denotes the estimated coefficient in the $i$-th row and $j$-th column of the autoregressive matrix obtained from the $m$-th replication; $\overline{\Phi}_{ij}$ represents the mean value of this coefficient across all $M$ replications and $\Phi_{ij,0}$ is its true population coefficient. To measure the estimated variance of the coefficient in the $i$-th row and $j$-th column of the autoregressive matrix, we use:
\begin{equation}
\label{eq:Var}
    \widehat{\text{Var}}(\hat{\Phi}_{ij}) = \frac{1}{M-1} \sum_{m=1}^{M} \left(\hat{\Phi}^{(m)}_{ij} - \overline{\Phi}_{ij} \right)^2. 
\end{equation}
The estimated MSE for each coefficient in the matrix $\Phi$ is given by: 
\begin{equation} 
\label{eq:MSE}
    \widehat{\text{MSE}}(\hat{\Phi}_{ij}) = \widehat{\text{Var}}(\hat{\Phi}_{ij}) + \left[\widehat{\text{Bias}}(\hat{\Phi}_{ij}) \right]^2,
\end{equation} 
where $\widehat{\text{Bias}}(\hat{\Phi}_{ij})$ and $\text{Var}(\hat{\Phi}_{ij})$ are defined in \eqref{eq:Bias} and \eqref{eq:Var}, respectively. Due to the high dimensionality of $\Phi$, reporting bias, variance, and MSE for each coefficient is impractical. Instead, we summarize the performance of the estimators—GCov, diagonal GCov, and RGCov—by averaging all the coefficients of the matrix obtained from \eqref{eq:Bias}, \eqref{eq:Var}, and \eqref{eq:MSE}. The results are based on $M = 1000$ replications with increasing sample sizes $T = (200, 500, 800)$, and are presented in Figure~\ref{fig:bias_variance_mseA} for the case $\delta_T = \delta$, and in Figure~\ref{fig:bias_variance_mseB_ss1} for the case $\delta_T = \eta/T$. Table~\ref{tab:summary1} summarizes the results for both cases.

We begin by analyzing the case $\delta_T = \delta$, focusing on the first row of Figure~\ref{fig:bias_variance_mseA}, which illustrates the performance of RGCov and GCov (a special case of RGCov when $\delta = 0$). In this three-dimensional plot, the $x$-axis denotes the sample size $T$, the $y$-axis represents the value of $\delta$, and the $z$-axis displays the performance metric (bias, variance, and MSE) computed by averaging the corresponding matrix entries, as described above. By examining increasing values of $\delta$, cross-validation can be applied to select the optimal $\delta$ that minimizes the respective metric. The results reveal a trade-off between the shrinkage coefficient and bias: while small values of $\delta$ (including $\delta = 0$) result in high bias due to insufficient regularization, large values of $\delta$ also increase bias in absolute terms by over-regularizing the estimator. However, between these extremes, for each $T$, there exists an optimal value of $\delta$ that minimizes the bias, indicated by gray cells in Table~\ref{tab:summary1}. This is not true for the variance and MSE. Indeed, Figure~\ref{fig:bias_variance_mseA} amd Table~\ref{tab:summary1} show that while large shrinkage coefficients introduce bias by distorting estimates, they also stabilize the estimator by reducing variance and MSE. In other words, excessive shrinkage introduces estimation bias, but it reduces variance and MSE. Finally, the diagonal RGCov results are presented in Table \ref{tab:summary1} and illustrated in the second row of Figure \ref{fig:bias_variance_mseA}. Each plot is two-dimensional—with $T$ on the $x$-axis and the analyzed metric on the $y$-axis—as $\delta$ is not involved in this estimation. The results highlight the better performance of the (R)GCov. Despite its computational simplicity, the diagonal GCov estimator loses crucial information by considering only the coefficients on the main diagonal of $\Gamma(0,\theta)$, making it less accurate and efficient than GCov and RGCov.

Let us now consider the case where the shrinkage coefficient decreases with the sample size, specifically the case $\delta_T = \eta / T$; see Table~\ref{tab:summary1} and Figure~\ref{fig:bias_variance_mseB_ss1}. In the latter, the sample size $T$ is shown on the $x$-axis, the parameter $\eta$ on the $y$-axis, and the performance metric on the $z$-axis. Although the shrinkage coefficient, $\delta_T$, is not directly represented as an axis, this visualization is preferable to a two-dimensional graph with $\delta_T$ on the $x$-axis, as it clearly shows how the metric varies with both $\eta$ and $T$. Furthermore, placing $T$ and $\eta$ on the $x$- and $y$-axes, respectively, allows us to evaluate not only their individual effects but also their combined effect through the ratio $\eta/T$, reflecting the shrinkage coefficient definition. It should be noted that for the GCov and diagonal GCov, the results are the same as the previous case since these estimators do not depend on the shrinkage coefficient.

The results reveal a positive relationship between the shrinkage coefficient and bias. In particular, a high shrinkage coefficient leads to a negative bias; for instance, $\delta_T(\eta = 800, T = 200)$. Conversely, as the shrinkage coefficient decreases, the estimation bias shifts in the opposite direction, increasing toward positive and larger values, as highlighted by the point $\delta_T(\eta = 200, T = 800)$. It should be noted that, as the sample size increases, a larger $\eta$ is required to minimize the bias, as indicated by the gray cells in Table~\ref{tab:summary1}, which mark the optimal $\eta$ for each $T$. This pattern underscores the importance of selecting a smaller $\eta$ in small samples to control bias. At the same time, larger samples require a larger $\eta$ to prevent $\delta_T$ from becoming too small and losing its regularization effect. As in the case of $\delta_T = \delta$, variance and MSE respond differently to shrinkage than bias: while large shrinkage coefficients introduce bias by distorting estimates, they also stabilize the estimator by reducing variance and MSE. 

Finally, Table~\ref{Tab:Iden_ss1} summarizes the frequency with which the correct mixed causal and noncausal model is identified—defined as having fourteen eigenvalues inside and one eigenvalue outside the unit circle—in the two simulation settings corresponding to $\delta_T = \delta$ and $\delta_T = \eta/T$. The results indicate that identification accuracy improves with larger sample sizes. Moreover, a relatively constant $\delta$ performs well in the first scenario, while higher values of $\eta$ become preferable as $T$ increases in the second scenario.

\begin{table}[H]
    \centering
    \caption{Bias, variance, and MSE for RGCov (under the settings $\delta_t = \delta$ and $\delta_t = \eta/T$), GCov, and diagonal GCov, computed across different values of $T$, $\delta$, and $\eta$. Each metric is calculated element-wise and summarized by averaging over all entries of the corresponding matrices. Cells highlighted in gray indicate the values of the shrinkage parameter ($\delta$ or $\eta$) that optimize each metric. The high dimensionality results from the number of variables $n$.}
\renewcommand{\arraystretch}{1.2}
\setlength{\tabcolsep}{4pt}
\resizebox{\textwidth}{!}{%
\begin{tabular}{c|cccc|cccc|ccc|ccc}
\toprule
\multirow{2}{*}{\textbf{T}} 
 & \multicolumn{4}{c|}{\textbf{RGcov: Case $\boldsymbol{\delta_t = \delta}$}} 
 & \multicolumn{4}{c|}{\textbf{RGcov: Case $\boldsymbol{\delta_t = \eta/T}$}} 
 & \multicolumn{3}{c|}{\textbf{GCov}} 
 & \multicolumn{3}{c}{\textbf{Diagonal GCov}} \\
\cmidrule(lr){2-5} \cmidrule(lr){6-9} \cmidrule(lr){10-12} \cmidrule(lr){13-15}
 & $\boldsymbol{\delta}$ & Bias & Var & MSE
 & $\boldsymbol{\eta}$    & Bias & Var & MSE 
 & Bias & Var & MSE
 & Bias & Var & MSE \\
\midrule


\multirow{9}{*}{\textbf{200}}
 & \textbf{0.5} & \cellcolor{gray!20}{-0.0016} & 0.09 & 0.10  & 
 \textbf{100} &  \cellcolor{gray!20}{-0.0016} & 0.09 & 0.10
 & -0.0024 & 0.97 & 0.97 
 & 0.0277 & 32.8 & 32.8 \\

 & \textbf{1.0} & -0.0019 & 0.04 & 0.04 & 
 \textbf{200} & -0.0019 & 0.04 & 0.04  
 & & & & & & \\

 & \textbf{1.5} & -0.0020 & 0.02 & \cellcolor{gray!20}{0.02} & 
 \textbf{300} & -0.0020 & 0.02 & \cellcolor{gray!20}{0.02}
 & & & & & & \\

 & \textbf{2.0} & -0.0020  & 0.02 & 0.02 & 
 \textbf{400} & -0.0020  & 0.02 & 0.02
 & & & & & & \\

 & \textbf{2.5} & -0.0019 &  \cellcolor{gray!20}{0.01} & 0.02 & 
 \textbf{500} &  -0.0019 & \cellcolor{gray!20}{0.01} & 0.02
 & & & & & & \\

 & \textbf{3.0} & -0.0019 & 0.01 & 0.02 &
 \textbf{600} &  -0.0019 & 0.01 & 0.02  
 & & & & & & \\

 & \textbf{3.5} & -0.0020 & 0.01 & 0.02 & 
 \textbf{700} & -0.0020 & 0.01 & 0.02
 & & & & & & \\

 & \textbf{4.0} & -0.0020 & 0.01 & 0.02 & 
 \textbf{800} &  -0.0020 & 0.01 & 0.02
 & & & & & & \\

\midrule


\multirow{9}{*}{\textbf{500}}
 & \textbf{0.5} &-0.0010 & 0.18 & 0.18 & 
 \textbf{100} & -0.0009 & 0.43 & 0.43 
 & 0.0022 & 0.91 & 0.91 & 
 0.0474 & 27.7 & 28.0 \\

 & \textbf{1.0} & \cellcolor{gray!20}{-0.0004} & 0.04 & 0.04 & 
 \textbf{200} & -0.0008 & 0.20 & 0.20 
 & & & & & & \\

 & \textbf{1.5} & -0.0010 & \cellcolor{gray!20}{0.01} &  \cellcolor{gray!20}{0.01} & 
 \textbf{300} & -0.0003 &  0.11 & 0.12 
 & & & & & & \\

 & \textbf{2.0} & -0.0011 & 0.02 & 0.02 & 
 \textbf{400} & -0.0006 & 0.07 & 0.07
 & & & & & & \\

 & \textbf{2.5} & -0.0012 & 0.01 & 0.01 & 
 \textbf{500} &  \cellcolor{gray!20}{-0.0004} & 0.04 & 0.04 
 & & & & & & \\

 & \textbf{3.0} & -0.0013 & 0.01 & 0.01 &
 \textbf{600} & -0.0010 & 0.02 & 0.02 
 & & & & & & \\

 & \textbf{3.5} & -0.0013 & 0.01 & 0.01 & 
 \textbf{700} & -0.0011 & 0.02 & 0.02
 & & & & & & \\

 & \textbf{4.0} & -0.0013 & 0.01 & 0.01 & 
 \textbf{800} & -0.0010 &  \cellcolor{gray!20}{0.01} &  \cellcolor{gray!20}{0.01}
 & & & & & & \\

\midrule


\multirow{9}{*}{\textbf{800}}
 & \textbf{0.5} & 0.0008 & 0.09 & 0.09 & 
 \textbf{100} & 0.0037 & 0.32 & 0.32 
 & 0.0037 & 0.46 & 0.46 
 & 0.0442 & 13.0 & 13.1 \\

 & \textbf{1.0} &  \cellcolor{gray!20}{0.0002} & 0.02 & 0.02 & 
 \textbf{200} & 0.0031 & 0.21 & 0.21 
 & & & & & & \\

 & \textbf{1.5} & -0.0007 &  \cellcolor{gray!20}{0.01} &  \cellcolor{gray!20}{0.01}& 
 \textbf{300} & 0.0017 &  0.17 & 0.17 
 & & & & & & \\

 & \textbf{2.0} & -0.0007 & 0.02 & 0.02 & 
 \textbf{400} & 0.0008 & 0.09 & 0.09 
 & & & & & & \\

 & \textbf{2.5} & -0.0009 & 0.01 & 0.01 & 
 \textbf{500} & 0.0009 & 0.10 & 0.10 
 & & & & & & \\

 & \textbf{3.0} & -0.0011 & 0.01 & 0.01 & 
 \textbf{600} & 0.0005 & 0.07 & 0.07 
 & & & & & & \\

 & \textbf{3.5} & -0.0011 & 0.01 & 0.01 & 
 \textbf{700} & 0.0003 & 0.05 & 0.05 
 & & & & & & \\

 & \textbf{4.0} & -0.0012 & 0.01 & 0.01 & 
 \textbf{800} &  \cellcolor{gray!20}{0.0002} & \cellcolor{gray!20}{0.02} & \cellcolor{gray!20}{0.02}
 & & & & & & \\

\bottomrule
\end{tabular}}
\label{tab:summary1}
\end{table}

\begin{table}[H]
\centering
\setlength{\tabcolsep}{8pt} 
\renewcommand{\arraystretch}{1.2} 
\caption{Identification frequencies (in $\%$) for the RGCov and diagonal Gcov (D. GCov) methods in correctly identifying the true process, characterized by one eigenvalue outside the unit circle and fourteen inside, for various sample sizes ($T$) and shrinkage coefficients ($\delta_T$). Results are presented for two cases: (1) $\delta_T = \delta$ and (2) $\delta_T = \frac{\eta}{T}$. Cells highlighted in gray indicate the values of $\delta$ and $\eta$ that maximize the percentage of correct identifications for each $T$. The high dimensionality results from the number of variables $n$.}
\label{Tab:Iden_ss1_extended}
\resizebox{\textwidth}{!}{%
\begin{tabular}{llccccccccc|c}
\toprule
$\boldsymbol{Case \ \delta_T \to \delta}$ &  & $\boldsymbol{\delta=0}$ & $\boldsymbol{\delta=0.5}$ & $\boldsymbol{\delta=1}$ & $\boldsymbol{\delta=1.5}$ & $\boldsymbol{\delta=2}$ & $\boldsymbol{\delta=2.5}$ & $\boldsymbol{\delta=3}$ & $\boldsymbol{\delta=3.5}$ & $\boldsymbol{\delta=4}$ & \textbf{D. GCov} \\
\midrule
\multirow{3}{*}{$\boldsymbol{\delta_T=\delta}$} 
    & $\boldsymbol{T=200}$  & 0.0 & 18.4 & \cellcolor{gray!20}\textbf{38.6} & 38.5 & 36.4 & 32.3 & 28.3 & 27.1 & 25.2 & 0.0 \\
    & $\boldsymbol{T=500}$  & 16.9 & 43.8 & 54.4 & \cellcolor{gray!20}\textbf{57.1} & 52.7 & 51.2 & 48.9 & 46.2 & 42.2 & 13.2 \\
    & $\boldsymbol{T=800}$  & 61.8 & \cellcolor{gray!20}\textbf{71.5} & 71.0 & 67.2 & 64.5 & 61.3 & 59.2 & 56.4 & 54.1 & 53.6 \\
\bottomrule
$\boldsymbol{Case \ \delta_T \to 0}$ &  & $\boldsymbol{\eta=0}$ & $\boldsymbol{\eta=100}$ & $\boldsymbol{\eta=200}$ & $\boldsymbol{\eta=300}$ & $\boldsymbol{\eta=400}$ & $\boldsymbol{\eta=500}$ & $\boldsymbol{\eta=600}$ & $\boldsymbol{\eta=700}$ & $\boldsymbol{\eta=800}$ & \textbf{} \\
\midrule
\multirow{3}{*}{$\boldsymbol{\delta_T=\frac{\eta}{T}}$} 
&$\boldsymbol{T=200}$	&	0.0	&	18.4	&	\cellcolor{gray!20}\textbf{38.6}	&	38.5	&	36.4	&	32.3	&	28.3	&	27.1	&	25.2 & 	\\
&$\boldsymbol{T=500}$	&	16.9	&	29.4	&	39.4	&	46.3	&	52.1	&	54.4	&	55.9	&	\cellcolor{gray!20}\textbf{57.8}	&	56.6 & 	\\
&$\boldsymbol{T=800}$	&	61.8	&	64.3	&	67.9	&	69.9	&	71.5	&	72.3	&	72.1	&	\cellcolor{gray!20}\textbf{72.4}	&	71.0 & 	\\
\bottomrule
\end{tabular}}
\end{table}

\subsection{High Dimensionality Driven by the Number of Transformations}
\label{subsec:large_J}
This section analyzes the performance of GCov, diagonal GCov, and RGCov in scenarios where the curse of dimensionality of the matrix $ \Gamma(0,\theta) $ is driven by $ J $, i.e., the number of transformations in \eqref{3.2}. The use of nonlinear transformations in the GCov is essential for correctly capturing the noncausal component and identifying the coefficients that result in an $i.i.d.$ error term [see \cite{gourieroux2023generalized}]. However, the choice of both the number and type of nonlinear transformations is not trivial and depends on the specific characteristics of the investigated data. In other words, a limited set of nonlinear transformations, or an inadequate number of transformations, may not be sufficient to capture the underlying features of the process. Such an approach reduces the need to carefully pre-select specific transformations, providing greater flexibility in modeling diverse data characteristics. Consequently, the GCov estimator becomes more adaptable across different datasets, improving its overall reliability [see \cite{cubadda2024optimization}]. However, while incorporating a large number of nonlinear transformations can improve the model’s ability to capture complex relationships, it also increases the dimensionality of the matrix $\Gamma(0,\theta)$, potentially leading to computational challenges and numerical instability. To address this issue, we investigate whether relying solely on the diagonal elements of $\Gamma(0,\theta)$ or applying the RGCov estimator to regularize the high-dimensional matrix can alleviate the adverse effects of dimensionality. This approach aims to balance the modeling flexibility—gained by including a rich set of transformations—with computational feasibility and stability.

We consider a 3-dimensional mixed VAR(1) model as DGP, characterized by two real eigenvalues inside the unit circle, $\lambda_1 = \{ 0.20, 0.41 \}$, and one real eigenvalue outside the unit circle, $ \lambda_2 = 1.5 $. As in the previous simulation study, we assume a multivariate Student-$t$ distribution with degrees of freedom $\nu = 4$ and a scale matrix equal to the identity matrix. For all three estimators, we set $H=2$ and $J=10$. In particular, we consider the following linear and nonlinear transformations: linear transformation $a_1\bigl[g_i(y_t; \theta)\bigr] = u_{i,t}$, quadratic transformation $a_2\bigl[g_i(y_t; \theta)\bigr] = u_{i,t}^2$, cubic transformation $a_3\bigl[g_i(y_t; \theta)\bigr] = u_{i,t}^3$, sign transformation $a_4\bigl[g_i(y_t; \theta)\bigr] = \text{sign}({u_{i,t}})$, absolute value transformation $ a_5\bigl[g_i(y_t; \theta)\bigr]= |u_{i,t}| $, cubic absolute transformation $ a_6\bigl[g_i(y_t; \theta)\bigr] = |u_{i,t}|^3 $, logarithmic transformation $ a_7\bigl[g_i(y_t; \theta)\bigr] = \log(|u_{i,t}|) $, squared logarithmic transformation $ a_8\bigl[g_i(y_t; \theta)\bigr] = \log(|u_{i,t}|)^2 $, cubic logarithmic transformation $a_9\bigl[g_i(y_t; \theta)\bigr] = \log(|u_{i,t}|)^3 $, and square root absolute transformation $ a_{10}\bigl[g_i(y_t; \theta)\bigr] = |u_{i,t}|^{1/2} $. These transformations are designed to capture a variety of different dynamic features in the data. The linear transformation preserves the original data structure, while the quadratic and cubic transformations capture higher-order dependencies typical in time series with volatility clustering. Transformations such as the absolute value and sign transformations are used to further model volatility and extreme fluctuations. In particular, these transformations separate dynamic volatility from other effects, such as bid-ask bounce, by highlighting the magnitude and directionality of fluctuations [see \cite{gourieroux2023generalized}]. The cubic absolute transformation emphasizes large values, which are crucial for capturing extreme events, while logarithmic transformations compress large values to handle diminishing returns and improve robustness to outliers. Finally, the square root absolute transformation helps reduce the impact of large values or extreme fluctuations in the data, making it useful for stabilizing variance.

\begin{figure}[H]
    \centering
    \caption{Bias, variance, and MSE of the (R)GCov estimator (first row) and the diagonal GCov estimator (second row) under the setting $\delta_T = \delta$, for varying values of $\delta$ and sample sizes $T$. Increasing values of $\delta$ are considered to facilitate cross-validation and identify the optimal shrinkage level that minimizes the respective performance metric. Each metric is computed element-wise and then averaged over all matrix entries. The high dimensionality in this setting arises from the number of transformations $J$.}
    \includegraphics[width=18.5cm, height=7cm]{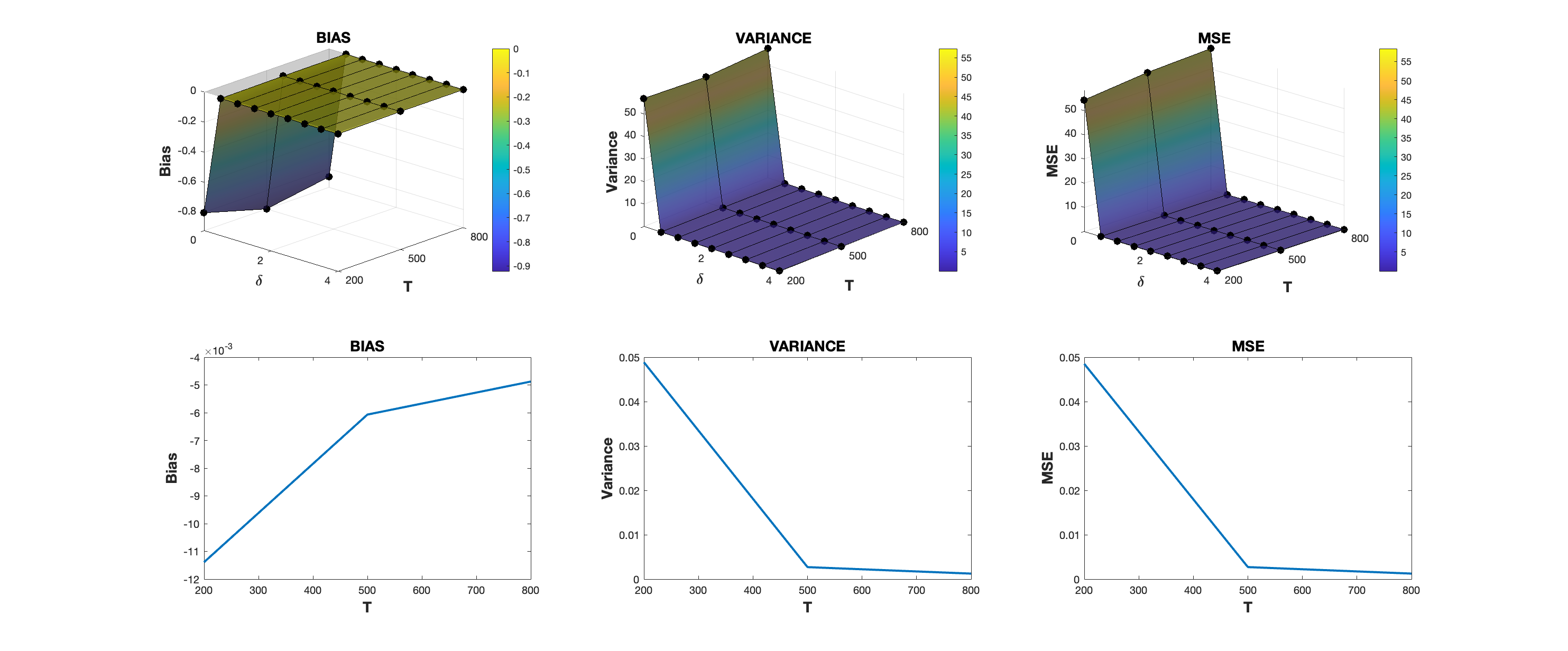}
    \label{fig:BiasVarianceA_ss2}
\end{figure}

\begin{figure}[H]
    \centering
    \caption{The graph shows the bias, variance, and MSE of the RGCov estimator under the setting $\delta_T = \eta/T$, for increasing values of $\eta$ and sample sizes $T$. Each metric is computed element-wise and then averaged over all matrix entries. The high dimensionality in this setting arises from the number of transformations $J$.}
    \includegraphics[width=18.5cm, height=3.3cm]{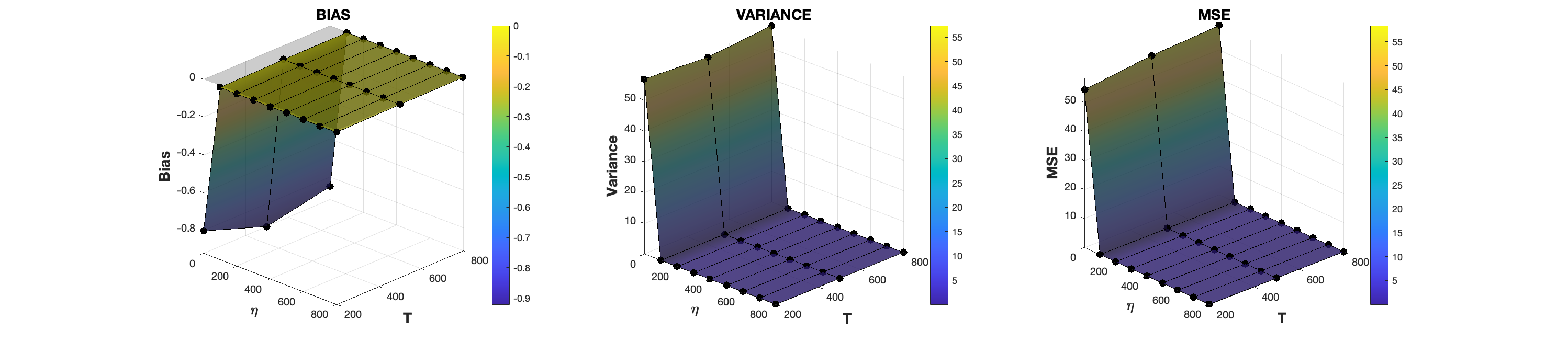}
    \label{fig:BiasVarianceB_ss2}
\end{figure}

Even in this case, we assess the performance of the three estimators using bias, estimated variance, and MSE, as defined in \eqref{eq:Bias}, \eqref{eq:Var}, and \eqref{eq:MSE}, respectively. In particular, as discussed in Section~\ref{subsec:large_n}, each metric is computed element-wise and then summarized by averaging over all entries of the corresponding matrix. Figure~\ref{fig:BiasVarianceA_ss2}, Figure~\ref{fig:BiasVarianceB_ss2}, report the results for the diagonal GCov and the (R)Gcov under the settings $ \delta_T=\delta$ and $ \delta_T=\eta/T$, respectively, while Table~\ref{tab:summary2} summarize the results for all cases. Table~\ref{Tab:Iden_ss2} presents the percentage of correct identifications of the true DGP, characterized by one eigenvalue outside the unit circle and two inside. 

The results confirm the bad performance of the GCov estimator and reveal two key differences compared to those discussed in Section~\ref{subsec:large_n}. First, the diagonal GCov performs well when the high dimensionality of the matrix $\Gamma(0,\theta)$ is driven by the number of nonlinear transformations $J$, particularly in terms of variance and MSE for larger sample sizes ($T = 500, 800$). Second, a small shrinkage coefficient is required for RGCov not only to minimize bias but also to jointly optimize both bias and MSE in both scenarios: $\delta_T = \delta$ and $\delta_T = \eta/T$.

\begin{table}[H]
\centering
\caption{Bias, variance, and MSE for RGCov (under the settings $\delta_t = \delta$ and $\delta_t = \eta/T$), GCov, and diagonal GCov, computed across different values of $T$, $\delta$, and $\eta$. Each metric is calculated element-wise and summarized by averaging over all entries of the corresponding matrices. Cells highlighted in gray indicate the values of $\eta$ that optimize each metric. The high dimensionality results from the number of transformations $J$.}
\renewcommand{\arraystretch}{1.2}
\setlength{\tabcolsep}{4pt}
\resizebox{\textwidth}{!}{%
\begin{tabular}{c|cccc|cccc|ccc|ccc}
\toprule
\multirow{2}{*}{\textbf{T}} 
 & \multicolumn{4}{c|}{\textbf{RGcov: Case $\boldsymbol{\delta_t = \delta}$}} 
 & \multicolumn{4}{c|}{\textbf{RGcov: Case $\boldsymbol{\delta_t = \eta/T}$}} 
 & \multicolumn{3}{c|}{\textbf{GCov}} 
 & \multicolumn{3}{c}{\textbf{Diagonal GCov}} \\
\cmidrule(lr){2-5} \cmidrule(lr){6-9} \cmidrule(lr){10-12} \cmidrule(lr){13-15}
 & $\boldsymbol{\delta}$ & Bias & Var & MSE
 & $\boldsymbol{\eta}$    & Bias & Var & MSE 
 & Bias & Var & MSE
 & Bias & Var & MSE \\
\midrule


\multirow{9}{*}{\textbf{200}}


 & \textbf{0.5} & -0.0089 & \cellcolor{gray!20}{0.01} & \cellcolor{gray!20}{0.01}  & 
 \textbf{100} & -0.0089 & \cellcolor{gray!20}{0.01} & \cellcolor{gray!20}{0.01} &  
 -0.8029 & 56.9 & 54.4 &
 -0.0114  & 0.05 & 0.05 \\

 & \textbf{1.0} & -0.0109 & 0.01 & 0.01& 
 \textbf{200} & -0.0109 &0.01 & 0.01 &
 & & &  &  &  \\

 & \textbf{1.5} & -0.0091 & 0.01 & 0.01  & 
 \textbf{300} & -0.0091 & 0.01 & 0.01  &
 &  &  &  &  &  \\

 & \textbf{2.0} & -0.0108 & 0.01 & 0.01  &
 \textbf{400} & -0.0108 & 0.01 & 0.01 &
 &  &  &  &  &  \\

 & \textbf{2.5} & \cellcolor{gray!20}{-0.0078}  & 0.01 & 0.01 & 
 \textbf{500} & \cellcolor{gray!20}{-0.0078} & 0.01 & 0.01 &
 &  & &  &  &  \\

 & \textbf{3.0} & -0.01 & 0.01 & 0.01 &
 \textbf{600} & -0.0108 &  0.01 & 0.01&  &
 & &  &  &  \\

 & \textbf{3.5} & -0.0114 & 0.01 &0.01  &
 \textbf{700} & -0.0114 & 0.01 &0.01 &
 &  &   &  &  &  \\

 & \textbf{4.0} & -0.0103 & 0.01 & 0.01 &
 \textbf{800} & -0.0103  & 0.01 & 0.01 & 
 &  & &  &  &  \\

\midrule


\multirow{9}{*}{\textbf{500}}


 & \textbf{0.5} & \cellcolor{gray!20}{-0.0045} & \cellcolor{gray!20}{ 0.01} &  \cellcolor{gray!20}{ 0.01}
 & \textbf{100} & \cellcolor{gray!20}{-0.0039} & \cellcolor{gray!20}{ 0.01} & \cellcolor{gray!20}{ 0.01} 
 &-0.9211  & 55.58 & 57.17 
 & -0.0061 & 0.0027 & 0.0027 \\

 & \textbf{1.0} & -0.0050 & 0.01 & 0.01 & 
 \textbf{200} &  -0.0042 &  0.01 &  0.01 &  &  &  &  &  &  \\

 & \textbf{1.5} & -0.0068 &  0.01 &  0.01 & \textbf{300} & -0.0044 & 0.01  &  0.01  &  &  & &  &  &  \\

 & \textbf{2.0} & -0.0047 &  0.01 &  0.01 & \textbf{400} & -0.0049 & 0.01& 0.01 &  &  &  &  &  &  \\

 & \textbf{2.5} & -0.0070 &  0.01 &  0.01 & \textbf{500} & -0.0050 &  0.01 &  0.01 &  &  & &  &  &  \\

 & \textbf{3.0} & -0.0068 & 0.01 & 0.01  &
 \textbf{600} & -0.0051 &  0.01 &  0.01 &  &  & &  &  &  \\

 & \textbf{3.5} &-0.0069 &  0.01 &  0.01 & 
 \textbf{700} & -0.0057 &  0.01 &  0.01 &  &  & &  &  &  \\

 & \textbf{4.0} & -0.0050 & 0.01 &  0.01
       & \textbf{800} & -0.0056 &  0.01 &  0.01
       &  &  & 
       &  &  &  \\

\midrule


\multirow{9}{*}{\textbf{800}}


 & \textbf{0.5} & \cellcolor{gray!20}{-0.0028} & \cellcolor{gray!20}{ 0.01} & \cellcolor{gray!20}{ 0.01}
       & \textbf{100} & -0.0029 & \cellcolor{gray!20}{0.01} & \cellcolor{gray!20}{0.01}
       & -0.8521 & 57.38 & 58.39
       & -0.0049 & 0.0012  & 0.0013 \\

 & \textbf{1.0} & -0.0039 &  0.01 & 0.01 
       & \textbf{200} & \cellcolor{gray!20}{-0.0025} & 0.01 & 0.01
       &  &  & 
       &  &  &  \\

 & \textbf{1.5} & -0.0034 & 0.02 & 0.02 
       & \textbf{300} & -0.0030 & 0.01 &  0.01
       &  &  & 
       &  &  &  \\

 & \textbf{2.0} & -0.0051 & 0.01 & 0.01 
       & \textbf{400} & -0.0028 & 0.01 & 0.01 
       &  &  & 
       &  &  &  \\

 & \textbf{2.5} & -0.0043 & 0.01 & 0.01
       & \textbf{500} & -0.0029 & 0.01 & 0.01
       &  &  & 
       &  &  &  \\

 & \textbf{3.0} & -0.0034& 0.01 &  0.01
       & \textbf{600} & -0.0030 &  0.01 & 0.01 
       &  &  & 
       &  &  &  \\

 & \textbf{3.5} & -0.0035 & 0.01 & 0.01
       & \textbf{700} & -0.0031 & 0.01 & 0.01
       &  &  & 
       &  &  &  \\

 & \textbf{4.0} & -0.0048 & 0.01 & 0.01
       & \textbf{800} & -0.0039 & 0.01 & 0.01
       &  &  & 
       &  &  &  \\

\bottomrule
\end{tabular}}
\label{tab:summary2}
\end{table}

\begin{table}[H]
\centering
\setlength{\tabcolsep}{9pt}
\renewcommand{\arraystretch}{1.25}
\caption{Identification frequencies (in $\%$) for the RGCov and diagonal Gcov (D. GCov) methods in correctly identifying the true process, characterized by one eigenvalue outside the unit circle and fourteen inside, for various sample sizes ($T$) and shrinkage coefficients ($\delta_T$). Results are presented for two cases: (1) $\delta_T = \delta$ and (2) $\delta_T = \frac{\eta}{T}$. Cells highlighted in gray indicate the values of $\delta$ and $\eta$ that maximize the percentage of correct identifications for each $T$. The high dimensionality arises from the number of transformations, $J$.}
\label{Tab:Iden_ss2}
\resizebox{\textwidth}{!}{%
\begin{tabular}{llccccccccc|c}
\toprule
\textbf{Case:} $\boldsymbol{\delta_T = \delta}$ & & $\boldsymbol{\delta=0}$ & $\boldsymbol{0.5}$ & $\boldsymbol{1}$ & $\boldsymbol{1.5}$ & $\boldsymbol{2}$ & $\boldsymbol{2.5}$ & $\boldsymbol{3}$ & $\boldsymbol{3.5}$ & $\boldsymbol{4}$ & \textbf{D. GCov} \\
\midrule
& $\boldsymbol{T = 200}$ & 12.3 & 85.9 & 86.0 & \cellcolor{gray!20}\textbf{86.8} & 85.7 & 85.2 & 85.3 & 85.3 & 86.8 & 90.5 \\
& $\boldsymbol{T = 500}$ & 11.2 & 92.5 & \cellcolor{gray!20}\textbf{93.3} & 91.6 & 90.6 & 91.1 & 91.0 & 91.8 & 91.9 & 97.2 \\
& $\boldsymbol{T = 800}$ & 9.4 & \cellcolor{gray!20}\textbf{97.3} & 97.2 & 97.3 & 96.6 & 96.9 & 96.4 & 96.0 & 95.5 & 95.1 \\
\midrule
\textbf{Case:} $\boldsymbol{\delta_T = \frac{\eta}{T}}$ & & $\boldsymbol{\eta=0}$ & $\boldsymbol{100}$ & $\boldsymbol{200}$ & $\boldsymbol{300}$ & $\boldsymbol{400}$ & $\boldsymbol{500}$ & $\boldsymbol{600}$ & $\boldsymbol{700}$ & $\boldsymbol{800}$ & \\
\midrule
& $\boldsymbol{T = 200}$ & 12.3 & 85.9 & 86.0 & \cellcolor{gray!20}\textbf{86.8} & 85.7 & 85.2 & 85.3 & 85.3 & 86.8 & \\
& $\boldsymbol{T = 500}$ & 11.2 & \cellcolor{gray!20}\textbf{93.6} & 93.5 & 93.5 & 93.5 & 93.3 & 92.5 & 92.4 & 93.5 & \\
& $\boldsymbol{T = 800}$ & 9.4 & \cellcolor{gray!20}\textbf{98.0} & 97.9 & 98.2 & 97.3 & 97.8 & 98.0 & 97.2 & 97.2 & \\
\bottomrule
\end{tabular}
}
\end{table}

\section{Application}

This section considers the monthly stock price series of 12 green energy companies included in the Renixx index. 
 The Renixx index tracks the global renewable energy market, covering sectors such as wind, solar, bioenergy, geothermal, hydropower, electronic mobility, and fuel cells. It comprises 30 companies, each of which derives more than 50\% of its revenues from these sectors (see \url{www.iwr.de/renixx}). Companies are selected based on the highest market capitalization of freely traded stocks (float market capitalization), ensuring that no single sector comprises more than 50\% of the index. Furthermore, the combined weight of the fuel cell, mobility, and utility companies is limited to no more than 20\% of the index, with no individual security exceeding a weight of 10$\%$. The Renixx is a total return performance index that reflects both price and dividend changes, quoted in euros. It is re-weighted quarterly and reviewed semi-annually, with adjustments made for new initial public offerings.

\begin{figure}[H]
    \centering
    \caption{ Scaled Prices for VAR Estimation}
    \includegraphics[width=12cm, height=8cm]{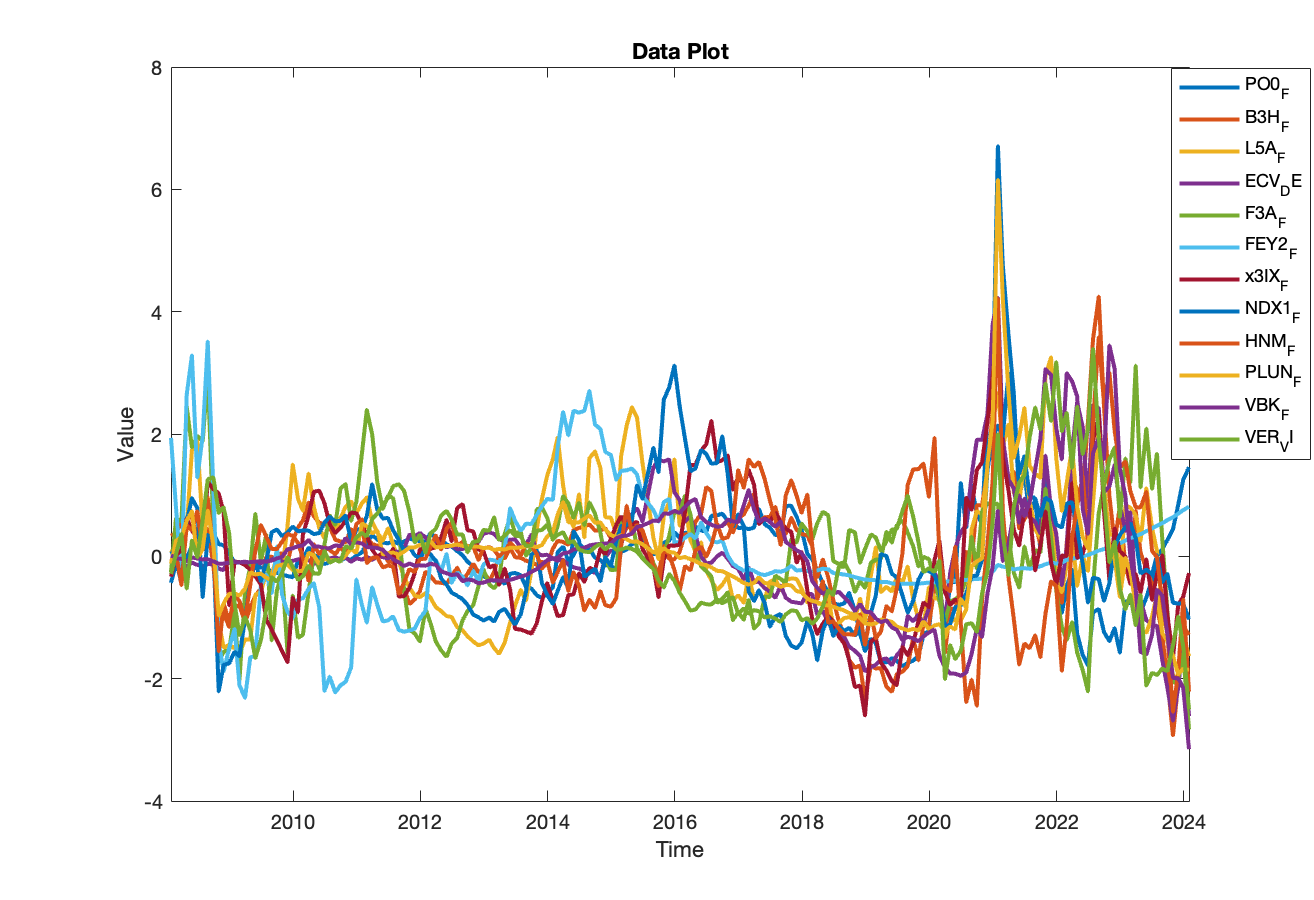}
    \label{App1}
\end{figure}

The sample we consider in this paper start in January 2008 and ends in January 2024, giving us a $T=192$ observations. We detrend each series by the cubic-spline method (see Hall and Jasiak, 2024) and divide them by their standard deviation. The plots of the price series are provided in Figure \ref{App1}.  We fit mixed-VAR(1) with both GCov and RGCov estimators and compare the results. We consider four transformations of residuals up to power four of them and a number of lags included in the GCov estimator $H=10$. Under these settings, the dimension of the covariance matrix is $48\times48$, which is evidence we are in the high dimensional framework. For the RGCov estimator, we consider fixed $\delta=0.1,\dots,  0.5$ and choose the one that gives eigenvalues far from the unit root. Table \ref{tab:application} demonstrates the eigenvalues of a lag polynomial of mixed-VAR(1) based on the GCov and RGCov estimators.

According to Table \ref{tab:application}, The GCov estimator gives eigenvalues close to one. Also, the value of the GCov test is significantly higher than the ones provided by RGCov. The estimation time is also more than twice that of RGCov. Another interesting fact is the estimates of mixed-VAR(1) based on GCov and RGCov provide different identification in terms of causal-noncausal orders. The GCov estimator provides 11 eigenvalues higher than one; however, RGCov with different regularization parameter values provides only one eigenvalue higher than one.  To choose the regularization parameter of RGCov, we compare the closest eigenvalues to the unity for each of them and choose the one with a higher distance, excluding $\delta=0.1$, since it gives an eigenvalue close to zero. Based on this approach, we choose $\delta=0.3$ for more comparison with GCov for the rest of this section.

\begin{table}[H]
    \centering
    \caption{Table of sorted modulus eigenvalues, test values, and time for different $\delta$ values.}
    \resizebox{\textwidth}{!}{%
    \begin{tabular}{lcccccc}
        \hline
         & Gcov & RGCov($\delta=0.1$) & RGCov($\delta=0.2$) & RGCov($\delta=0.3$) & RGCov($\delta=0.4$) & RGCov($\delta=0.5$) \\
        \hline
        \textbf{Sorted eigenvalues} \\
        & 0.91 & \textbf{0.05} & 0.24 & 0.29 & 0.29 & 0.29 \\
        & 0.91 & 0.27 & 0.24 & 0.29 & 0.29 & 0.29 \\
        & \textbf{0.98} & 0.68 & 0.56 & 0.39 & 0.41 & 0.42 \\
        & \textbf{1.02} & 0.68 & 0.64 & 0.70 & 0.70 & 0.70 \\
        & \textbf{1.02} & 0.76 & 0.64 & 0.70 & 0.70 & 0.70 \\
        & 1.10 & 0.76 & 0.81 & 0.79 & 0.78 & 0.77 \\
        & 1.10 & 0.88 & 0.86 & 0.79 & 0.78 & 0.77 \\
        & 1.50 & 0.88 & 0.86 & 0.90 & 0.93 & 0.94 \\
        & 2.25 & 0.89 & 0.89 & 0.95 & 0.95 & 0.96 \\
        & 2.25 & 0.89 & 0.89 & 0.96 & 0.97 & 0.96 \\
        & 4.88 & 0.90 & \textbf{1.00} & \textbf{0.96} & \textbf{0.97} & \textbf{0.98 }\\
        & 4.88 & \textbf{1.06} & 1.08 & 1.09 & 1.08 & 1.08 \\
        \hline
        \textbf{Test value} & 20729.67 & 4873.19 & 3060.46 & 2207.95 & 1709.69 & 1385.13 \\
        \hline
        \textbf{Time} & 34.52 & 15.13 & 13.26 & 11.67 & 11.75 & 9.39 \\
        \hline
    \end{tabular}
    }
    \label{tab:application}
\end{table}

\begin{figure}[H]
\caption{ Residuals}
    \centering
     \begin{subfigure}[b]{0.49\textwidth}
         \centering
         \includegraphics[width=\textwidth, height = 6cm]{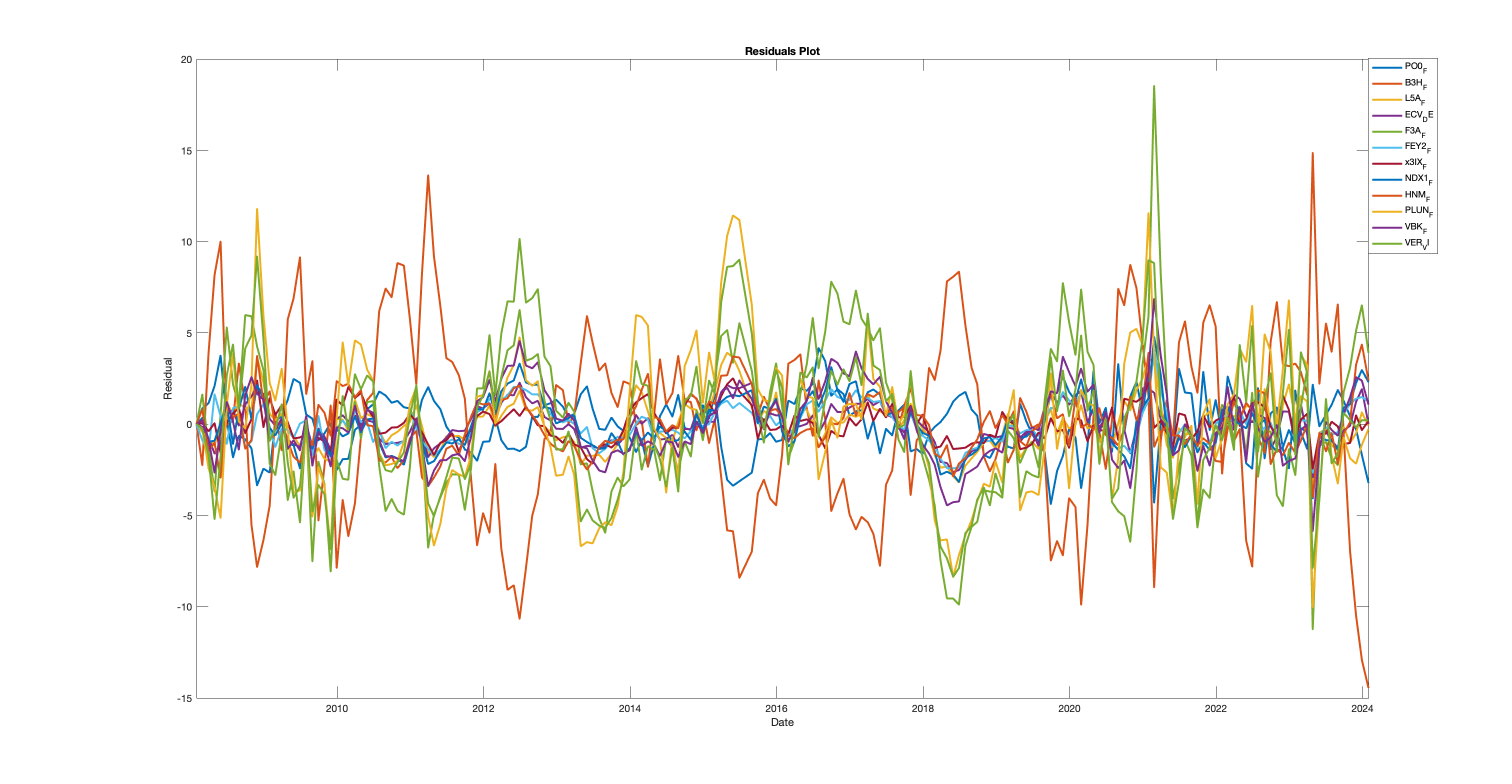}
         \caption{GCov}
         \label{app2-a}
     \end{subfigure}
     \hfill
         \begin{subfigure}[b]{0.49\textwidth}
         \centering
         \includegraphics[width=\textwidth, height = 6cm]{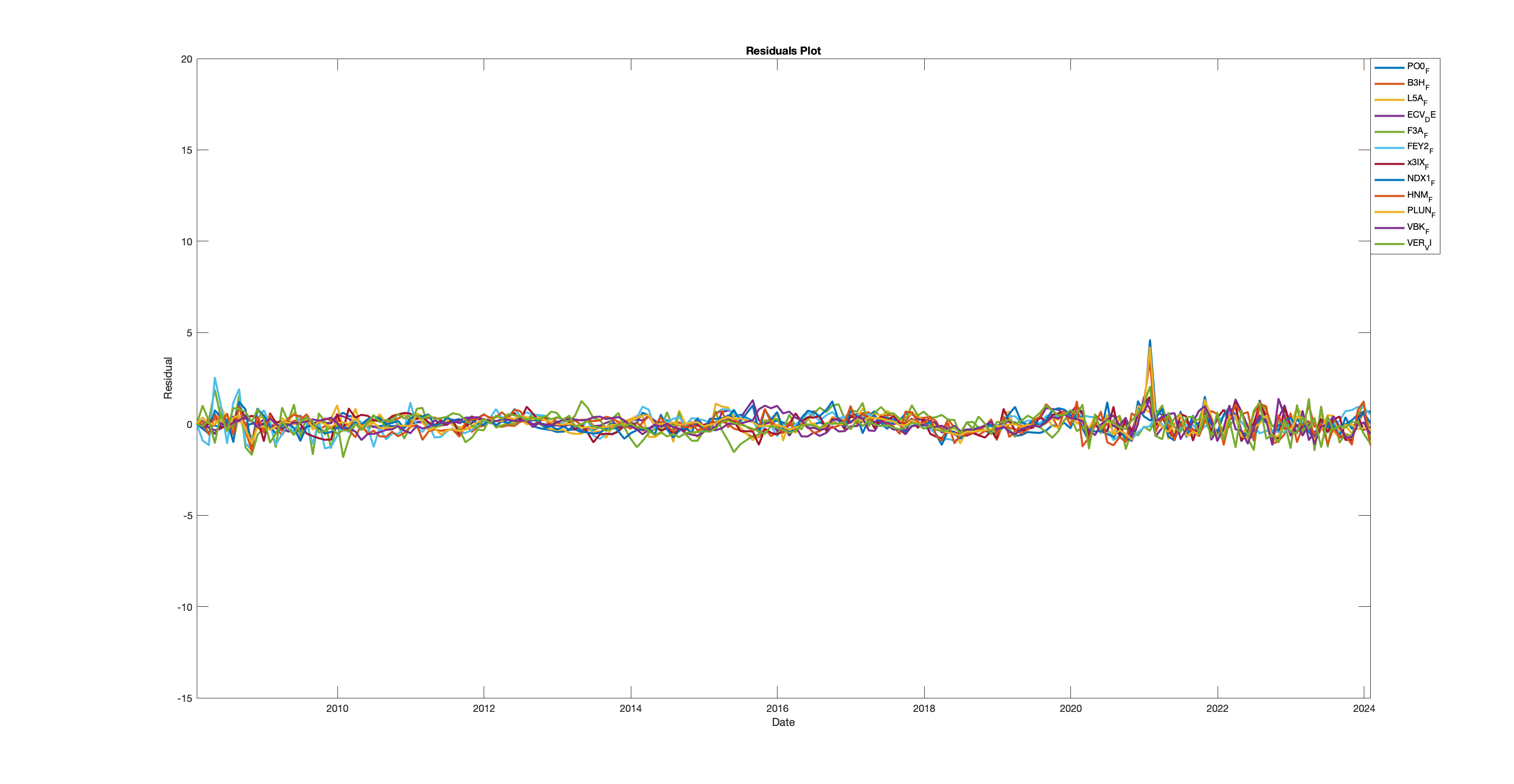}
         \caption{RGCov}
         \label{app2-b}
     \end{subfigure}
     \label{app2}
\end{figure}

The GCov estimator in the high-dimensional setting has an invertibility issue for $\hat{\Gamma}(0)$, and using RGCov solves that problem. Table \ref{tab:3} provides evidence for this fact by reporting the ten lowest eigenvalues of $\hat{\Gamma}(0)$ and $\hat{\Gamma}(0,\delta)$ for $\delta=0.3$. Let us go deeper to compare the results of GCov and RGCov with $\delta=0.3$. Figure \ref{app2} shows the estimated residuals of each of the estimators. The residuals of RGCov(\ref{app2-b}) show a much more steady pattern than the residuals from GCov(\ref{app2-a}). 

\begin{table}[H]
    \centering
    \caption{The ten lowest eigenvalues of $\hat{\Gamma}(0)$ and $\hat{\Gamma}(0,\delta)$ for $\delta=0.3$.}
    \resizebox{\textwidth}{!}{%
    \begin{tabular}{lcccccccccc}
    \hline
        & \multicolumn{10}{c}{\textbf{Sorted eigenvalues}} \\
        \hline
         \textbf{GCov}  & 0.0 & 0.0 & 0.0 & 0.0 & 0.0 & 0.0 & 0.0001 & 0.0004 & 0.0006 & 0.0012  \\
         \textbf{RGCov} & 0.0006 & 0.0740 & 0.2952 & 0.6434 & 0.8501 & 1.1095 & 1.2171
 & 1.2737 & 1.5902 & 1.6305 \\
        \hline
    \end{tabular}
    }
    
    \label{tab:3}
\end{table}

\begin{figure}[H]
\caption{ Causal and noncausal components of VAR(1) estimated by RGCov $\delta=0.3$}
    \centering
     \begin{subfigure}[b]{0.49\textwidth}
         \centering
         \includegraphics[width=\textwidth, height = 6cm]{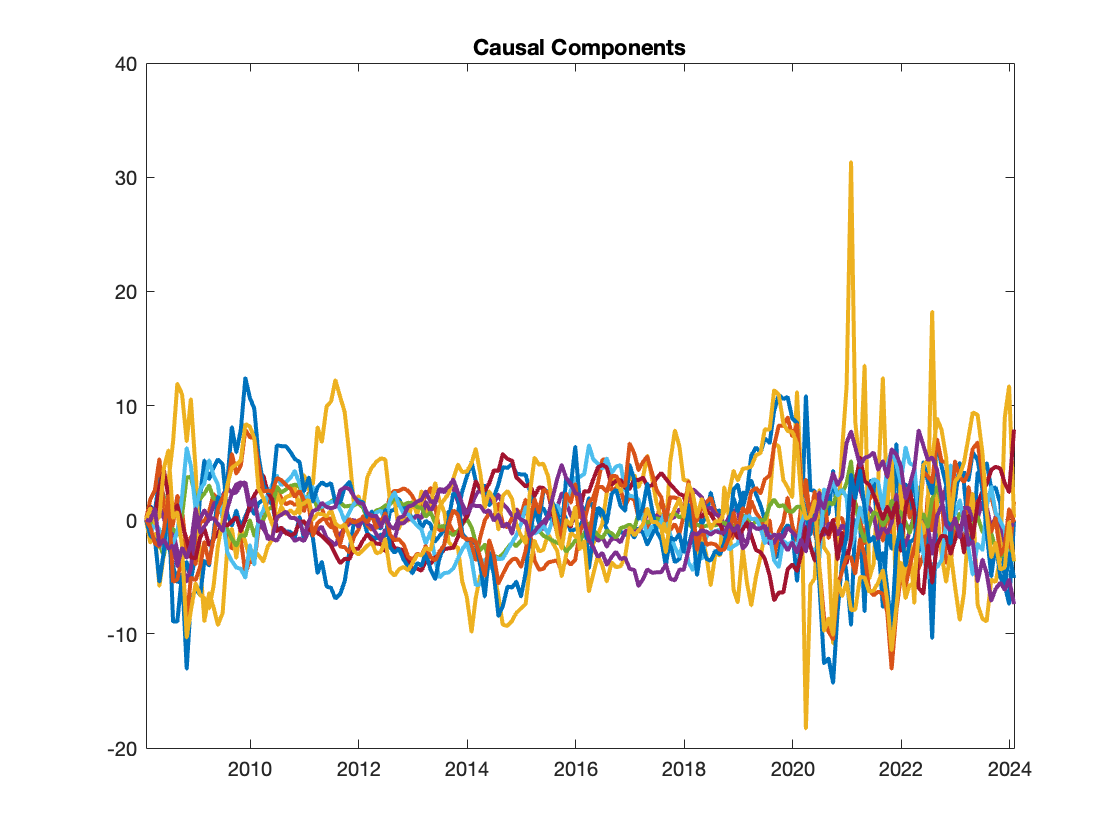}
         \caption{Causal components }
         \label{app2-a}
     \end{subfigure}
     \hfill
         \begin{subfigure}[b]{0.49\textwidth}
         \centering
         \includegraphics[width=\textwidth, height = 6cm]{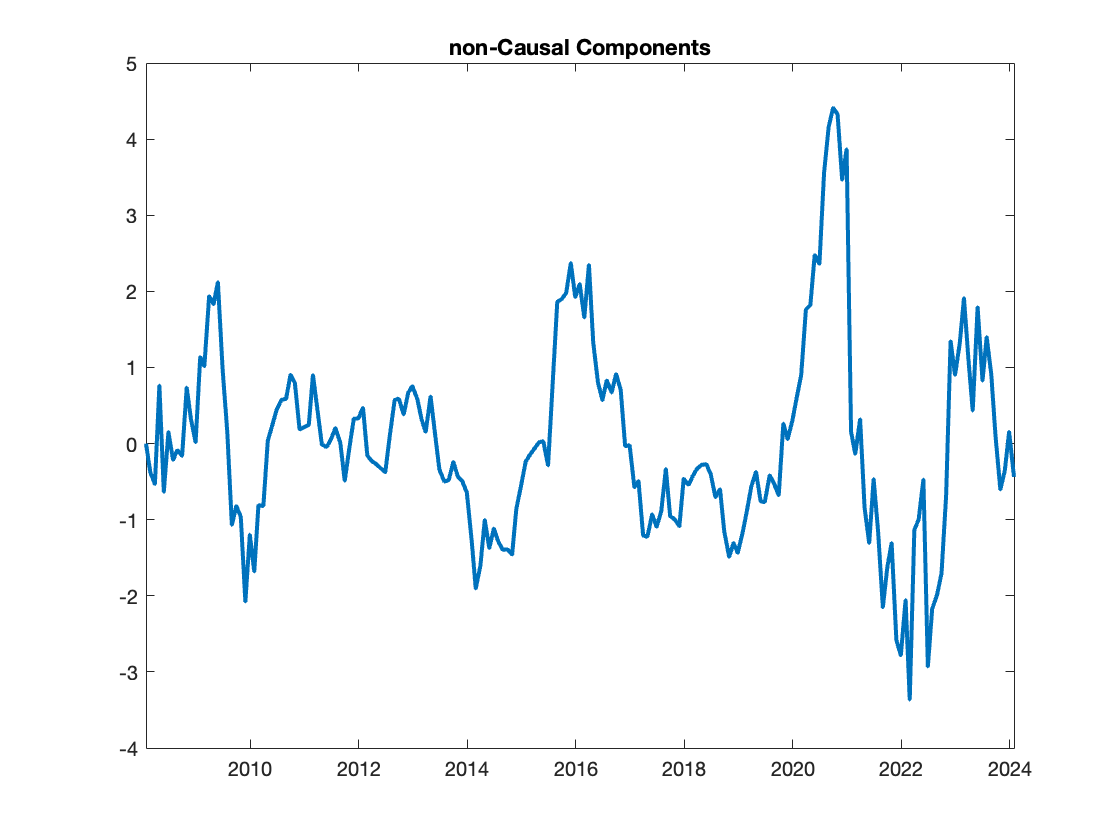}
         \caption{Noncausal components}
         \label{app2-b}
     \end{subfigure}
     \label{app3}
\end{figure}

\indent The representation theorem introduced by \cite{gourieroux2017noncausal} for mixed processes distinguishes between their purely causal and noncausal latent components. For the mixed VAR($1$) model with a diagonalizable autoregressive coefficient matrix\footnote{Otherwise, we use the real Jordan form of matrix $\Phi$.} we can rewrite the matrix $\bPhi$ as:
\begin{equation*}
    \bPhi=\bA \bJ \bA^{-1},
\end{equation*}
where $\bJ$ is a diagonal matrix with eigenvalues of $\bPhi$ on the diagonal, and $\bA$ is an invertible matrix consisting of the eigenvectors of  $\bPhi$. Furthermore, let us denote by $n_1$  the number of eigenvalues of $\bPhi$ with a modulus strictly less than 1 and by $n_2$ the number of eigenvalues with a modulus strictly greater than 1, where $n_2 = n - n_1$. Then, it is possible to express $\bJ$ as follows:
\begin{equation*}
    \bJ=\begin{bmatrix}
        \bJ_1 & \boldsymbol{0} \\ 
        \boldsymbol{0} & \bJ_2 \end{bmatrix},
\end{equation*}
where $\bJ_1$ is of dimension $n_1 \times n_1$ and has on its main diagonal the eigenvalues of $\bPhi$ that lie inside the unit circle, while $\bJ_2$ is of dimension $n_2 \times n_2$
and has on its diagonal the eigenvalues of $\bPhi$ that lie outside the unit circle. Consequently:
\begin{align}
& \by_t = \bA_1 \by^*_{1,t} + \bA_2 \by^*_{2,t}, \\
& \by^*_{1,t} = \bJ_1 \by^*_{1,t-1} + \bveps^*_{1,t}, \quad
 \by^*_{2,t} = \bJ_2 \by^*_{2,t-1} + \bveps^*_{2,t},\label{eq:causNoncaus} \\
& \bveps^*_{1,t} = \bA^1 \bveps_t, \quad
 \bveps^*_{2,t} = \bA^2 \bveps_t, \\
& \by^*_{1,t} = \bA^1 \by_t, \quad
 \by^*_{2,t} = \bA^2 \by_t, \label{eq:linear}
\end{align}
\noindent where $\bA_1, \bA_2$ represent the blocks in the decomposition of matrix $A$ as~:
$\bA=(\bA_1, \bA_2)$, and $\bA^1, \bA^2$ represent the blocks in the decomposition of $\bA^{-1}$ as $\bA^{-1} = \left(\bA_1, \bA_2\right)^{\prime}$.\\
\indent Since the eigenvalues of $\bJ_{1}$ in (\ref{eq:causNoncaus}) are in modulus less than 1, $\by_{1,t}^*$ is defined as the latent purely causal component of $\by_t$. By similar reasoning, $\by^*_{2,t}$ captures the noncausal component of the investigated process. In particular, $\by^*_{2,t}$ is the locally explosive component of $\by_t$ that follows a strictly stationary noncausal (V)AR process. Figure \ref{app3} illustrates the causal and noncausal components of VAR(1) estimated with RGCov estimator with $\delta=0,3$.

As highlighted in \cite{hall2024modelling} and following \cite{engle1987co}, equation (\ref{eq:linear}) suggests that the causal and noncausal components are represented by linear combinations that eliminate the noncausal and causal components, respectively. Specifically, in a mixed VAR($1$) process, we have $n_1 + n_2 = n$, which implies $n_1 < n$ and $n_2 < n$, hence indicating the presence of common causal and noncausal components by definition. Conversely, if the process is either purely causal or purely noncausal, then $n = n_1$ or $n = n_2$, respectively. This means that if the process is either purely causal or purely noncausal, each series in the investigated data is driven by independent causal or noncausal components, assuming that $\bA$ is a full-rank matrix. Hence, the causal component $\by^*_{1,t}$ is the stationary linear combination that eliminates its local explosive characteristics and can be interpreted as bubble 'cointegration' in the sense that it removes the nonlinear patterns captured by the noncausal component [see \cite{hall2024modelling}].\footnote{Note that \cite{cubadda2023detecting} and \cite{cubadda2019detecting} investigate the presence of co-movements in the multiplicative VAR model, necessitating the choice of models in terms of lag and lead polynomials.} In the next subsection, we propose portfolio management based on the causal and noncausal components of the VAR(1) estimated by RGCov with $\delta=0.3$.

\subsection{Portfolio Management}

In the framework of causal-noncausal VAR models, bubbles arise as inherent features of a strictly stationary multivariate process. The bubbles are common to the component series, estimable, and predictable because the bubble dynamics are approximated by the non-causal component of the VAR process. In addition, the linear combinations of green stocks that eliminate common bubbles follow directly from the state-space representation of the causal-noncausal VAR model and are estimable and predictable as well from the causal components of the VAR. Hence, the causal and non-causal components, including the bubble, can be interpreted as ”common features” of the green stocks in the sense of \cite{engle1993testing} and compared with the bubble cointegration of \cite{cubadda2024optimization}. The causal component of the VAR model is a bubble-free linear combination of green stocks. It represents stable investment portfolios, while the portfolios that "ride the bubble" can be obtained from the noncausal combinations of green stocks. The latter portfolios assume more risk but provide higher returns during the bubble episode.

Let us consider the causal and the noncausal components of the mixed VAR(1) fitted to twelve green stocks in the Renixx index. We investigate the performance of portfolios of these stocks with the allocations determined from the coefficients of the causal and the noncausal components of the process. The objective is to compare the performance of the Rennix index with the stable portfolio and the portfolio with the allocations determined from the explosive noncausal component in terms of cumulative returns. To construct the portfolios, we focus our attention on the current causal component $A^1_t$ and noncausal component $A^2_t$.

We consider an investor who invests $V_1=100\$$ in the green stocks at time $t=1$ and sells the stocks at the end of the month for the amount of $V_2=V_1+r_2$, where $r_2$ is the return over the first month. At time  $t=2$, s/he invests $V_2=V_1+r_2$. Therefore, at any time $t=1,..T$, the investor invests $V_t=V_1 + \sum_{h=2}^t r_h$ and sells the stocks at the end of each month. This strategy is repeated for $t=1,2,..., T-1$. The negative coefficients in the allocation vector provided by the causal and non-causal components are interpreted as a short sell. The initial investment of $V_1$ is equal for all of the portfolios.

We can construct 12 causal and noncausal, i.e. stable portfolios from the allocations given in $A^1_t$ and $A^2_t$. We consider $a_{ij}$ indicates $i^{th}$ portfolio and $j^{th}$ asset. According to the budget constraints on $V_t$ in each period, the weights are equal to $w_{ijt}=s_{it}* a_{ij}$ where $s_{it} $ is:

$$s_{it}= \frac{V_{it}}{\sum_{j=1}^{12} |a_{ij}|*p_{j,t}},$$

\nin  where $V_{it}$ is the value of the investment at time $t$ in portfolio $i$ and $p_{jt}$ is the price of asset j at time t. To estimate the return of each portfolio, we have
$$r_{i,t+1}= \sum_{j=1}^{12} w_{ijt}*p_{j{t+1}} - \sum_{j=1}^{12} w_{ijt}*p_{j{t}}.$$
$$r_{i,t+1}= s_{it} \sum_{j=1}^{12}  b_{ij} [( ln(p_{j,{t+1}}) -ln(p_{j,{t}})].$$
\nin Consequently, the cumulative sums of returns are the summation of the returns over time. 

 Since Renixx is an index of green stocks with positive weights, we calculate the returns on Rennix as 
$$r_{Renixx,t+1}= s_{Renixx,t}*[( ln(p_{Renixx,{t+1}}) -ln(p_{Renixx,{t}})],$$
\nin where 
$$s_{Renixx,t}=\frac{V_{it}}{p_{Renixx,{t}}}.$$

\begin{figure}[H]
    \centering
    \caption{Cumulated return }
    \includegraphics[width=12cm, height=8cm]{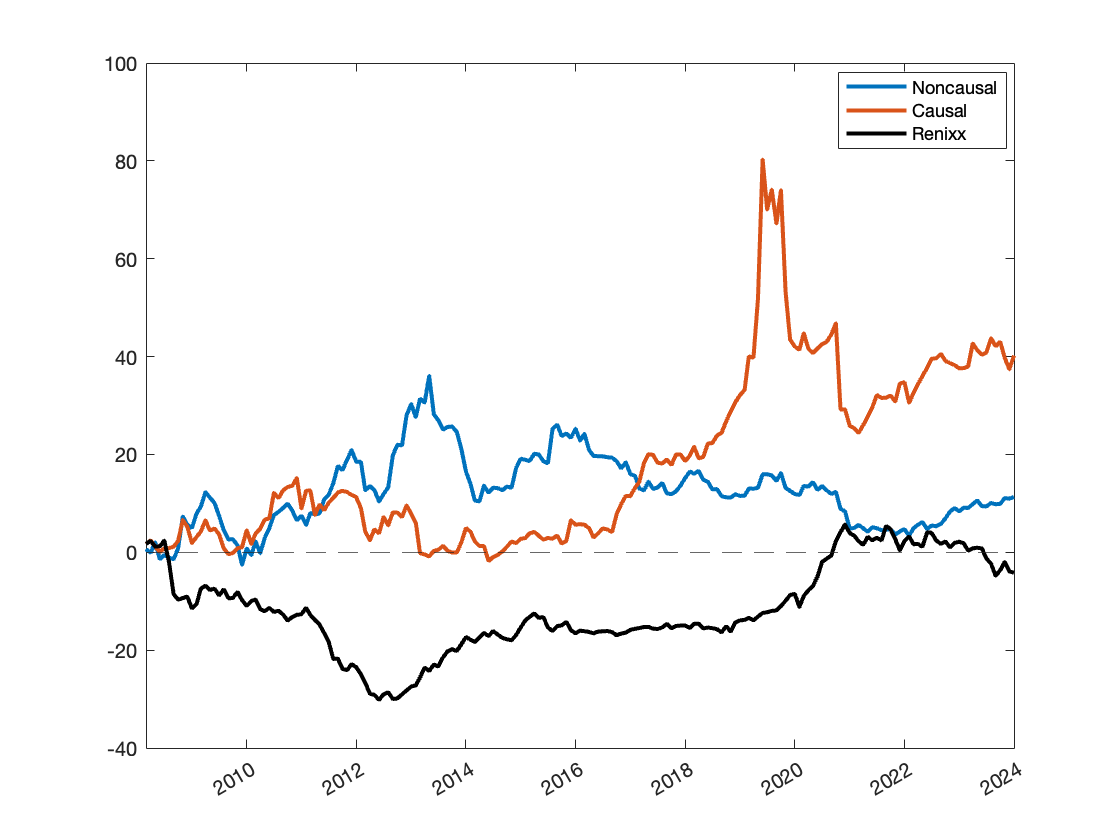}
    \label{app4}
\end{figure}

In Figure \ref{app4}, we provide the cumulated return of the noncausal portfolio and the best causal portfolio among 11 causal portfolios and the Renixx itself. Both proposed portfolios always eliminate Renixx in terms of cumulated returns. During the bubble period, the causal portfolio performs better. However, the noncausal portfolio with the bubble effect performance is steadier overall.

\section{Conclusion}

This paper proposes a regularized GCov estimator to address the invertibility issue of the variance matrix in the objective function of the GCov estimator due to the existence of many variables or consideration of many nonlinear transformations. We consider the Ridge-type regularization for the sample variance matrix, and we show that RGCov is consistent and has an asymptotically Normal distribution. Moreover, we provide conditions for the RGCov estimator to reach the semi-parametric efficiency bound. 

We introduce the RGCov specification test on estimated residuals and the RNLSD test to test the absence of linear and nonlinear serial dependence in time series. We show that both tests, based on the regularization parameter converging to a nonzero constant, have an asymptotically a mixture of chi-square distributions and, based on the regularization parameter converging to zero, have an asymptotically a chi-square distribution with known degrees of freedom. 

We explore the finite sample properties of the RGCov estimator based on vast simulation studies. We consider many variables and transformations that could cause an invertibility issue of the variance matrix. In both cases, the RGCov estimator performs significantly better than the GCov and diagonal GCov estimators. For an empirical illustration, we use the RGCov estimator to fit a mixed VAR model with roots inside and outside of the unit circle to 12 stock price series of green energy companies included in the Renixx index. Consequently, based on causal and noncausal components of mixed VAR(1), we proposed portfolios that perform better than the Renixx index in terms of cumulative rate of return.

\section*{Appendix A}

\setcounter{equation}{0}\def\theequation{a.\arabic{equation}}

\textbf{Proof of Proposition 1}

\medskip

\nin 1) \textbf{Proof of Consistency}

The demonstration follows the lines of \cite{gourieroux2023generalized}. 
We consider the case $\delta_T \rightarrow \delta \geq 0$, that includes both cases $\delta_T = \delta, \forall T$, and $\delta_T \rightarrow 0$.

\medskip
\nin Let us consider the following assumptions:

\nin \textbf{Assumption A.1}

i) The parameter set $\Theta$ is compact;

ii) The model is well-specified, with $\theta_0$ the true value of the parameter;

iii) $\theta_0$ is asymptotically identifiable, i.e. $\Gamma(h; \theta) = 0, \; h=1,...,H$ implies $\theta=\theta_0$; 

iv) The transformation $g$ is continuous in $\theta$, and the transformations $a_j, j=1,...,J$
are continuous in $u$,

v) The process $Y_t$ and errors $u_t(\theta_0) = g(\tilde{Y}_t; \theta_0)$, $t=1,2,...$  are strictly stationary, geometrically ergodic with a continuous invariant distribution,  

vi) $\delta_T$, $\delta_T>0 \; \forall T$,   tends to $\delta\geq 0$, when $T$ tends to infinity and $\Gamma(0, \theta_0, \delta)$ is invertible.

\medskip
\nin \textbf{Propositions 1 i)}:

Let $\hat{\theta}_T(\delta_T)$ denote a solution of the minimization:

$$\hat{\theta}_T(\delta_T) = Arg min_{\theta} \sum_{h=1}^H Tr R^2_T(h, \theta, \delta_T),$$

\nin where: 

$$R^2_T(h, \theta, \delta_T) = \hat{\Gamma}_T(h, \theta) [\delta_T I + \hat{\Gamma}_T(0, \theta)]^{-1} \hat{\Gamma}_T(h; \theta)'[\delta_T I + \hat{\Gamma}_T(0, \theta)]^{-1},$$

\nin then there exists such a solution for $T$ sufficiently large, such that $\hat{\theta}_T(\delta_T)$ tends to $\theta_0$, when $T$ tends to infinity.

\medskip

Proof: The result is obtained by applying the standard Jennrich argument. The objective function tends asymptotically uniformly to the limiting function:

$$\sum_{h=1}^H Tr R^2(h; \theta, \delta),$$

\nin where $R^2(h; \theta) = \Gamma(h; \theta) \Gamma(0; \theta, \delta)^{-1} \Gamma(h; \theta)' \Gamma(0; \theta, \delta)^{-1}$, and $\Gamma(h; \theta)$ are the autocovariances computed under the true distribution of $(Y_t)$. Moreover, the convergence is uniform in $\theta$. It follows that $\hat{\theta}_T(\delta_T)$ exists asymptotically and it converges to the solution of the asymptotic optimization. This is the solution $\theta$ of $\Gamma(h, \theta) = 0, h=1,...,H$, i.e. $\theta_0$ by the identification assumption. \hfill QED

\medskip
Therefore, the consistency does not depend on the limiting value of the regularization parameter.

\bigskip

\nin 2) \textbf{Proof of Asymptotic Normality}

The following additional assumptions are required:

\nin \textbf{Assumption A.2}

i) $\Theta$ has a non-empty interior $ \stackrel{o}{\Theta}$;

ii) $\theta_0 \in  \stackrel{o}{\Theta}$;

iii) The functions $g(\tilde{Y}, \theta)$ and the functions $a_j(u), j=1,...,J$, are twice continously differentiable with respect to $\theta$ and u, respectively.

iv) The transformations $g_j (\tilde{Y}_t; \theta), \; j=1,...,J$ satisfy the following equicontinuity condition:
$$sup_{j=1,...,J} | g_j^2 (\tilde{Y}_t; \theta) - g_j^2 (\tilde{Y}_t; \tilde{\theta})| \leq {\cal B}(\tilde{Y}_t) h [d(\theta, \tilde{\theta})],
\theta, \tilde{\theta} \in \Theta,$$
\nin where $d( \theta, \tilde{\theta})$ is the distance on the metric parameter space, $h(d)$ is a function that tends to 0 when $d$ tends to 0, and ${\cal B}(\tilde{Y}_t)$ is a sequence of non-negative variables, such that \\
$ sup_T \frac{1}{T} \sum_{t=1}^T E[{\cal B}(\tilde{Y}_t)] < \infty $.
This equicontinuity condition corresponds to Assumption W\_LIP in Andrews (1992), p 248, and is a stochastic Lipschitz condition on $g_j(Y; \theta)$.

\medskip

\nin The derivation of the asymptotic normality, i.e., of Proposition 1 ii),  proceeds in several steps, following \cite{gourieroux2023generalized}, Section 4.2.
\medskip

\nin b) First-order derivative of the theoretical objective function 

\nin For ease of exposition, we  consider lag the $h=1$ and write
the first-order derivative with respect to $\theta$ of T$r \, R^2 (1; \theta, \delta)$:

$$Tr \,R^2 (1;\theta, \delta) = Tr [\Gamma(1;\theta) \Gamma(0; \theta, \delta)^{-1} \Gamma(1; \theta)' \Gamma(0, \theta, \delta)^{-1}].$$

\nin where $\Gamma(0, \theta, \delta) = \delta I + \Gamma(0, \theta)$.

\nin The first-order partial derivatives are computed by considering the differential:
$$d \, Tr \, R^2(1; \theta, \delta) = \sum_{j=1}^J \frac{\partial Tr \, R^2 (1;\theta, \delta)}{\partial \theta_j} d\theta_j,$$

\nin which is found along the lines of \cite{gourieroux2023generalized}, Section 4.2 and Supplementary Material A.2.1:

\begin{eqnarray}
d \, Tr \,R^2(1; \theta, \delta) & = & 2 \,Tr \,[\Gamma(0; \theta, \delta)^{-1} \Gamma(1; \theta)' \Gamma(0; \theta, \delta)^{-1} d \Gamma(1; \theta)] \nonumber\\
& - & Tr \, \left[ \Gamma(0; \theta, \delta)^{-1} [ \Gamma(1; \theta)' \Gamma(0; \theta, \delta)^{-1} \Gamma(1; \theta) \right. \nonumber \\
& & + \Gamma(1; \theta) \Gamma(0; \theta, \delta)^{-1} \Gamma(1; \theta)'] \Gamma(0; \theta, \delta)^{-1} d \Gamma(0; \theta, \delta) \left. \right] \nonumber\\
& = & 2 Tr\, [ \Gamma(0; \theta, \delta)^{-1} \Gamma(1; \theta)' \Gamma(0; \theta, \delta)^{-1}  d \Gamma(1; \theta)] \nonumber\\
& - & Tr \, \left[ R^2(1; \theta, \delta) \Gamma(0; \theta, \delta)^{-1} + \Gamma(0; \theta, \delta)^{-1}
R^2(1; \theta, \delta)] d \Gamma(0; \theta, \delta) \right], 
\end{eqnarray}

\nin Let us denote:

$$L_T(\theta, \delta) = \sum_{h=1}^H Tr R_T^2(h, \theta, \delta),$$

\nin for ease of exposition.

\medskip
c) First-order conditions (FOC)

\nin The first-order conditions of the minimization are:

$$\frac{\partial L_T(\hat{\theta}_T, \delta_T)}{\partial \theta_j} = \sum_{h=1}^H \frac{ \partial Tr \, R_T^2(h; \hat{\theta}_{T}, \delta_T) }{ \partial \theta_j} = 0, \; j=1,...,J=dim \theta. $$

\nin They can be re-written by using the formula (a.1) above for any $h$. We get:

\begin{eqnarray}
&  & \sum_{h=1}^H \{ 2 \,Tr [ \hat{\Gamma}_T(0; \hat{\theta}_T, \delta_T)^{-1} \hat{\Gamma}_T(h; \hat{\theta}_T)' \hat{\Gamma}_T(0; \hat{\theta}_T, \delta_T)^{-1} 
\frac{\partial \hat{\Gamma}_T(h; \hat{\theta}_T)}{\partial \theta_j}] \nonumber\\
&  & - Tr \{ \,[R_T^2(h; \hat{\theta}_T, \delta_T) \hat{\Gamma}_T(0; \hat{\theta}_T, \delta_T)^{-1} +\hat{\Gamma}_T(0; \hat{\theta}_T, \delta_T)^{-1} R_T^2(h, \hat{\theta}_T, \delta_T)] \frac{\partial \hat{\Gamma}_T(0; \hat{\theta}_T, \delta_T)}{\partial \theta_j} \} \}= 0, \nonumber \\
& & \; j=1,...,J=dim \theta.
\end{eqnarray}

\nin where $\hat{\theta}_T = \hat{\theta}_T(\delta_T)$.

\medskip
d) Expansion of the FOC

\nin Let us now perform an expansion of the FOC in a neighborhod of the limiting values $\hat{\theta}_T \approx \theta_0$ and $\delta_T \approx \delta$. We get:

$$
\sqrt{T} \frac{\partial L_T (\theta_0, \delta)}{\partial \theta} = \frac{\partial^2 L_T (\theta_0, \delta)}{\partial \theta \partial \theta'} \sqrt{T} (\hat{\theta}_T - \theta_0) + \sqrt{T}  \frac{\partial^2 L_T (\theta_0, \delta)}{\partial \theta \partial \delta'} (\delta_T -\delta)+ o_p(1),
$$

\nin where $o_p(1)$ is negligible in probability. Next, we get:

$$\sqrt{T} (\hat{\theta}_T -\theta_0) = \left(-\frac{\partial^2 L_T(\theta_0,\delta)}{\partial \theta \partial \theta'} \right)^{-1} \sqrt{T} \frac{\partial L_T (\theta_0,\delta)}{\partial \theta} -  ( \frac{\partial^2 L_T(\theta_0,\delta)}{\partial \theta \partial \theta'} )^{-1} (\sqrt{T}\frac{\partial^2 L_T(\theta_0,\delta)}{\partial \theta \partial \delta'}) (\delta_{T} - \delta) + o_p(1)$$

The above expansion takes into account the orders of the different derivatives. We know that the estimated autocovariances $\hat{\Gamma}_T(h, \theta)$ and variances $\hat{\Gamma}_T(0, \theta, \delta)$  are consistent of $\Gamma(h, \theta_0)$ and
$\Gamma(0, \theta_0, \delta)$. Moreover, the autocovariances $\hat{\Gamma}_T(h, \theta_0)$ tend to zero at speed $1/\sqrt{T}$ and are asymptotically normally distributed. Thus, the decomposition is written to get the convergence to an invertible limit of $\frac{\partial^2 L_T(\theta_0, \delta)}{\partial \theta \partial \theta'}$, and convergence in distribution of $\sqrt{T} \frac{\partial L_T(\theta_0, \delta)}{\partial \theta}$ and $\sqrt{T} \frac{\partial^2  L_T(\theta_0, \delta)}{\partial \theta \partial \theta'}$ to a Gaussian vector and matrix, respectively.

In particular, we observe that the second term of the right-hand side is negligible if $\delta_T \rightarrow \delta$. Therefore, this expansion is equivalent to:

$$\sqrt{T} (\hat{\theta}_T-\theta_0) = \left(- \frac{\partial^2 L_T(\theta_0,\delta)}{\partial \theta \partial \theta'} \right)^{-1} \sqrt{T} \frac{\partial L_T (\theta_0,\delta)}{\partial \theta} + o_p(1).$$

\nin We deduce that:

$$ \sqrt{T} (\hat{\theta}_T - \theta_0) \stackrel{d}{\rightarrow} N(0, J(\theta_0, \delta)^{-1} I(\theta_0, \delta)J(\theta_0, \delta)^{-1}),$$

\nin where $J(\theta_0, \delta) = \lim_{T\rightarrow \infty} \frac{\partial^2 L_T(\theta_0, \delta)}{\partial \theta \partial \theta'}$ and $I(\theta_0,\delta) = Var \left[ \sqrt{T} \frac{\partial L_T(\theta_0, \delta)}{\partial \theta}\right]$.

\medskip

\nin When $\delta =0$, the limiting behaviour is the same as for the GCov estimator, which has been derived in \cite{gourieroux2023generalized}. We have:

$$I(\theta_0,0)=J(\theta_0,0),$$

\nin and
$$ \sqrt{T} (\hat{\theta}_T - \theta_0) \stackrel{d}{\rightarrow} N(0, J(\theta_0, 0)^{-1}),$$

\nin where $J(\theta_0, 0)  = 2  \sum_{h=1}^H \left\{ \frac{\partial vec \Gamma(h; \theta_0)'}{\partial \theta} [\Gamma (0; \theta_0, 0)^{-1}  \otimes   
\Gamma (0; \theta_0, 0)^{-1} ] \frac{\partial vec \Gamma(h; \theta_0)}{\partial \theta'} \right\} = J (\theta_0). $ 

\nin For other values of $\delta$, the expressions of matrices $I(\theta_0,\delta)$ and  $J(\theta_0,\delta)$ have to be derived.

\medskip
e) Expressions of matrices $J(\theta_0, \delta)$ and $I(\theta_0, \delta)$ when $\delta_T \rightarrow \delta>0$.

i) Following the proof in \cite{gourieroux2023generalized}, Appendix A.2.2 of Supplementary Material, we can compute the second-order derivative of the objective function at the limiting values and deduce that:

$$-\frac{\partial^2 L_T(\theta_0, \delta)}{\partial \theta \partial \theta'} \stackrel{T \rightarrow \infty}{\rightarrow} J(\theta_0, \delta),$$

\nin with 

$$
J (\theta_0, \delta)  =  2 \sum_{h=1}^H \left\{ \frac{\partial vec \Gamma(h; \theta_0)'}{\partial \theta} [\Gamma (0; \theta_0, \delta)^{-1}  \otimes   
\Gamma (0; \theta_0, \delta)^{-1} ] \frac{\partial vec \Gamma(h; \theta_0)}{\partial \theta'} \right\},
$$

\nin Therefore we have:

$$\sqrt{T} (\hat{\theta}_T- \theta_0) = [J(\theta_0, \delta)]^{-1} \sqrt{T} \frac{\partial L_T(\theta_0, \delta)}{\partial \theta} + o_p(1).$$

\medskip
ii) Let us now focus on the limiting score. Since at the limit $R^2(h, \theta_0, \delta)=0, h=1,...,H$, it follows that asymptotically:

\begin{eqnarray*}
\sqrt{T} \frac{\partial L_T(\theta_0, \delta)}{\partial \theta_j} & = & 2  \sum_{h=1}^H
Tr \, \left[\Gamma (0; \theta_0, \delta)^{-1} \sqrt{T} \hat{\Gamma}_T (h; \theta_0)' \Gamma (0; \theta_0, \delta)^{-1} \frac{\partial \Gamma(h; \theta_0)}{\partial \theta_j} \right] + o_p(1)\\
& = & 2  \sum_{h=1}^H
Tr \, \left[\Gamma (0; \theta_0, \delta)^{-1}  \frac{\partial \Gamma(h; \theta_0)}{\partial \theta_j} \Gamma (0; \theta_0, \delta)^{-1} \sqrt{T} \hat{\Gamma}_T (h; \theta_0) \right]+ o_p(1)\\
& = & 2  \sum_{h=1}^H
Tr \, \left\{ \frac{\partial vec \Gamma(h; \theta_0)}{\partial \theta_j}' \Gamma (0; \theta_0, \delta)^{-1}  \otimes  \Gamma (0; \theta_0, \delta)^{-1} vec [\sqrt{T} \hat{\Gamma}_T (h; \theta_0)] \right\}+ o_p(1).
\end{eqnarray*}

We know that under the serial independence condition, the vectors $vec[ \sqrt{T} \hat{\Gamma}_T(h, \theta_0)]$,
$h=1,...,H$, are asymptotically independent and have the same Gaussian distribution $N[ 0, \Gamma(0, \theta_0) \otimes \Gamma(0, \theta_0)]$ [\cite{chitturi1976distribution}, \cite{hannan1976asymptotic})]:

$$vec[\sqrt{T} \hat{\Gamma}'_T(h, \theta_0)] \sim N[0, \Gamma(0, \theta_0) \otimes \Gamma(0, \theta_0)],\; h=1,...,H$$

\nin Therefore $\sqrt{T} \frac{\partial L_T(\theta_0, \delta)}{\partial \theta} $ is asymptotically normally distributed, with mean zero and asymptotic variance:

\begin{eqnarray}
V_{asy}[\sqrt{T} \frac{\partial L_T(\theta_0, \delta)}{\partial \theta} ] & = & 4 \sum_{h=1}^H \left\{ \frac{\partial vec \Gamma(h; \theta)'}{\partial \theta} \left[\Gamma(0; \theta_0, \delta)^{-1} \otimes \Gamma(0; \theta_0, \delta)^{-1}\right] [\Gamma(0;\theta_0) \otimes \Gamma(0;\theta_0)] \nonumber\right.\\
& & \left. \left[\Gamma(0; \theta_0, \delta)^{-1} \otimes \Gamma(0; \theta_0, \delta)^{-1}\right]  \frac{\partial vec \Gamma(h; \theta)}{\partial \theta'} \right\} \nonumber \\
& = & I(\theta_0, \delta).
\end{eqnarray}

\bigskip

\nin\textbf{Proof of Proposition 3} 

For ease of exposition, we consider the case of the Regularized NLSD (RNLSD) test, i.e. $dim \theta=0$. The proof of the general case is similar, but has to account for the effect of the estimation step of $\theta$ that creates dependence between the elements of $Tr \hat{R}_T^2(h, \hat{\theta}_T(\delta_T), \delta_T), \, h=1,...,H$ [see, Gourieroux, Jasiak (2023) for the proof with $\delta_T=0$].

\nin If $dim \theta=0$, we have :

$$
\hat{\xi}_T(H, \delta_T) =  T \sum_{h=1}^H Tr \hat{R}_T^2(h, \delta_T).
$$

\nin Asymptotically, the test statistic is equivalent to:

\begin{eqnarray*}
\hat{\xi}_T(H, \delta_T) & \approx & \sum_{h=1}^H Tr [ \Gamma(0; \delta)^{-1} \sqrt{T} \hat{\Gamma}_T (h) \Gamma(0;  \delta)^{-1}  \sqrt{T} \hat{\Gamma}'_T(h)]\\
& = & \sum_{h=1}^H \left\{ \sqrt{T} vec \hat{\Gamma}_T'(h) [ \Gamma(0; \delta)^{-1} \otimes 
\Gamma(0;\delta)^{-1} ] \sqrt{T} vec \hat{\Gamma}_T(h) \right\} \\
& =& \sum_{h=1}^H [X_T(h)' \Omega X_T(h)],
\end{eqnarray*}

\nin where $X_T(h) = \sqrt{T} vec \hat{\Gamma}_T'(h)$ and $\Omega =  \Gamma(0; \delta)^{-1} \otimes 
\Gamma(0; \delta)^{-1} = [ \Gamma(0, \delta) \otimes 
\Gamma(0; \delta)]^{-1}$. Under the independence hypothesis, the variables $X_T(h), h=1,...,H$ are independent and follow asymptotically a normal distribution:

$$X_T(h) \sim N(0, \Sigma),$$

\nin where:

$$\Sigma =  \Gamma(0) \otimes 
\Gamma(0).$$

\nin For any $h=1,...,H$ the variable $X_T(h)' \Omega X_T(h)$ is not chi-square distributed because $\Omega \Sigma \Omega \neq \Omega $, unless 
$\Gamma(0) = \Gamma(0,\delta)$, where the last equality holds in the special case $\delta=0$
[Gourieroux, Monfort (1989), Corollary R.201]. 
To find the asymptotic distribution of the RNLSD test statistic, let us consider 
$\tilde{X}_T(h) = \Sigma^{-1/2} \sqrt{T} vec \hat{\Gamma}_T(h)$, such that asymptotically $\tilde{X}_T(h) \sim N(0, I)$. We get:

$$\hat{\xi}_T(H, \delta_T)  \approx \sum_{h=1}^H \left(  \tilde{X}_T'(h) \Sigma^{1/2} \Omega \Sigma^{1/2} \tilde{X}_T(h) \right).$$

\nin The symmetric positive definite matrix $\Sigma^{1/2} \Omega \Sigma^{1/2}$ can be written as

$$\Sigma^{1/2} \Omega \Sigma^{1/2} = Q \Lambda Q',$$

\nin where $Q$ is an orthogonal matrix of eigenvectors and $\Lambda$ is a diagonal matrix of eigenvalues denoted by $\lambda_{l}^*, \, l=1,...,K^2$.

\nin Let us denote $Z_T(h)= Q'\tilde{X}_T(h)$. These vectors are asymptotically independent multivariate standard normal, and we have:

\begin{eqnarray*}
 \hat{\xi}_T(H, \delta_T) & \stackrel{d}{\rightarrow} & \sum_{h=1}^H \sum_{l=1}^{K^2}
 [ \lambda_{l}^* Z_{l,T}(h)^2]  = \sum_{l=1}^{K^2} [ \lambda_{l}^* ( \sum_{h=1}^H 
 Z_{l,T}(h)^2)] \\
 & \equiv &  \sum_{l=1}^{K^2} [\lambda_{l}^* Z_l^*]
\end{eqnarray*}

\nin where the variables $ Z_l^*, \, l=1,...,K^2$ are independent, with the $\chi^2(H)$ distribution.

\newpage
\bibliographystyle{chicago}
\bibliography{refrence}

\end{document}